 \documentclass[final,5p,times,twocolumn,authoryear]{elsarticle}

\usepackage{amssymb}
\usepackage{lipsum}
\usepackage{amsmath}
\usepackage{graphicx}
\usepackage{caption}
\usepackage{subfigure}
\usepackage{float}
\usepackage{xcolor}
\usepackage{fmtcount}
\usepackage{hyperref}
\usepackage{dcolumn}
\usepackage{bm}

\journal{Physics Letters B}

\begin{document}

\begin{frontmatter}



\title{Taxonomy of periodic orbits and gravitational waves in a non-rotating Destounis-Suvorov-Kokkotas black hole spacetime}

\author[a]{Zhutong Hua}
\ead{huazhutong19@mails.ucas.ac.cn}
\author[a]{Zhen-Tao He}
\author[a]{Jing-Qi Lai}
\author[b]{Jiageng Jiao}
\ead{jiaojiageng@ucas.ac.cn}
\author[a,c]{and Yu Tian}
\ead{ytian@ucas.ac.cn}

\affiliation[a]{School of Physical Sciences, University of Chinese Academy of Sciences, Beijing 100049, China}
\affiliation[b]{International Centre for Theoretical Physics Asia-Pacific, University of Chinese Academy of Sciences, 100190 Beijing, China}
\affiliation[c]{Institute of Theoretical Physics, Chinese Academy of Sciences, Beijing 100190, China}

\begin{abstract}
    In this paper, we investigate periodic orbits of test particles around a non-rotating Destounis-Suvorov-Kokkotas black hole and the resulting gravitational waves.
    Firstly, we examine the properties of circular orbits and find that circular orbits could disappear when the deformation is large enough. 
    Then, using an orbital taxonomy, we characterize various periodic orbits with a triplet of integers, which describes the zoom-whirl behaviours.
    We also calculate the gravitational waveform signals generated by different periodic orbits, revealing the influence of the deformation on the gravitational wave, which can be potentially picked up by future space-based detectors.
\end{abstract}




\end{frontmatter}




\section{Introduction}
\label{introduction}
Since the successful detection of the gravitational-wave signal GW150914 in 2016 by LIGO and Virgo, gravitational waves have become a new messenger of cosmic exploration\citep{LIGOScientific:2016emj, LIGOScientific:2016aoc,  LIGOScientific:2016vbw, LIGOScientific:2016vlm}. These oscillations of spacetime were predicted by Einstein's theory of general relativity and have significantly broadened the research horizon on the physics frontier\citep{Barack:2018yly, Berti:2015itd}. Among all the sources for space-based gravitational wave detectors——such as Taiji\citep{Hu:2017mde}, Tianqin\citep{TianQin:2015yph, Gong:2021gvw}, and LISA\citep{Schutz:1999xj, amaro2017laser, Gair:2004iv}——the Extreme Mass Ratio Inspiral system (EMRI) is a crucial class of gravitational wave sources in the mHz frequency band, in which a stellar-mass object inspirals around a supermassive black hole\citep{Hughes:2000ssa}.
To characterize EMRI orbits, the study of particle trajectories beyond circular orbits around black holes is of great significance\citep{Cardoso:2020iji}. It helps to understand the dynamics around black holes, providing a powerful theoretical framework for addressing challenges in astrodynamics. Among all the particle trajectories, periodic orbits are particularly important, as all eccentric orbits will undergo approximately periodic orbits during their inspiral towards the central black hole\citep{Levin:2008mq, Levin:2009sk}. 
Periodic orbits exhibit topological characteristics of so-called "multi-leaf clovers" which means whirling around the central black hole and zooming out in quasi elliptical orbits \citep{Barack:2003fp, Glampedakis:2002ya}.
Three integers index orbits topologically. They represent scaling, rotation and vertex behavior, respectively\citep{Levin:2008mq}. Therefore, an orbital taxonomy can associate periodic orbits with rational numbers. Using this kind of classification, a series of discussions have already been carried out in different spacetimes, such as Schwarzschild and Kerr\citep{Levin:2008mq, Levin:2009sk, Compere:2021bkk}, Reissner-Nordström\citep{PhysRevD.82.083001}, quantum-corrected black hole\citep{Guo:2025scs, Yang:2024lmj}, Einstein-Æther black hole\citep{Lu:2025cxx}, and so on\citep{Liu:2018vea,Wei:2019zdf,Gao:2020wjz,Lin:2022wda,Wang:2022tfo,Lin:2023rmo,Tu:2023xab,Jiang:2024cpe,Zahra:2025tdo, Deng:2025wzz, Lin:2021noq, Deng:2020hxw, Melis:2025iaw, Gong:2025mne, Ahmed:2025azu, Sharipov:2025yfw,Alloqulov:2025ucf}. An effective way to display these orbits is to calculate gravitational waveforms, and in some cases waveforms have been studied before, see Refs.\citep{Lu:2025cxx, Yang:2024lmj, Zahra:2025tdo, Deng:2025wzz, Gong:2025mne, Ahmed:2025azu, Ahmed:2025shr} and references therein.

Periodic orbits not only play an important role in orbit taxonomy but also have significance in studying transient resonances. 
As mentioned before, generic bound orbits can be approximately described by periodic orbits with perturbation\citep{Levin:2008mq}. 
In Kerr spacetime, when radiation reaction is considered, small compact objects will go through low order periodic orbits and transient resonances as a post adiabatic effect will emerge, accompanied by mutations of the motion constants \citep{Flanagan:2010cd, Berry:2016bit}. 
Moreover, resonances are focused more often in non-Kerr spacetime, where the Carter symmetry is broken and the spacetime becomes non-integrable.
In this case, the resonance islands will significantly generate within the phase space\citep{Pan:2023wau, Gair:2007kr, Apostolatos:2009vu, Lukes-Gerakopoulos:2010ipp, Contopoulos:2011dz, Amaro-Seoane:2011rdr, Kiuchi:2004bv, Zelenka:2019nyp, Destounis:2023gpw, Destounis:2023khj, Destounis:2023cim, Destounis:2025tjn, Destounis:2021rko, Eleni:2024fgs, Lukes-Gerakopoulos:2021ybx}.

Among various studies on non-Kerr spacetime, a metric, developed by Destounis, Suvorov and Kokkotas\citep{PhysRevD.102.064041} (DSK), includes generic deviations from the Kerr spacetime with a parameter $\alpha_{Q}$ controlling the Carter symmetry. 
In their work, gravitational-wave glitches associated with $r$-$\theta$ resonant orbits have been studied\citep{Destounis:2021mqv,PhysRevD.102.064041}. However, a systematic classification of periodic orbits in this spacetime remains to be studied. 
Given the complexity of non-Kerr spacetime, comprehensive analysis of periodic orbits in the generic case of this kind of spacetime is a big challenge. Therefore, we explore the $r$-$\varphi$ periodic orbits of this spacetime in static limit. Under this circumstance, the null circular orbit can be analytically solved. Actually, we solve null and time-like circular orbits in this spacetime and find that the circular orbits will be qualitatively different from those in the Schwarzschild spacetime in some situation. We also give a relatively complete taxonomy of periodic orbits with a triplet $(z, w, v)$ describing the number of orbital branches, the orbits whirling and the movement of the small object between apastrons, respectively. Based on this taxonomy, we study how the deformation affects the shape of them. At last, we also calculate the gravitational waveform signals generated by different periodic orbits, analyze the effects of deformation on them and show the differences from Schwarzschild spacetime by calculating the mismatch, which can be used to test the DSK model in future space-based observation.

The paper is constructed as follows. 
In section \ref{sec2}, a brief introduction of the deformed spacetime is presented. 
After this, in section \ref{sec3} the details of different circular orbits are analyzed separately. 
Firstly, the equations of motion are derived for the test particle and the effective potential from the Lagrangian is written. Then we investigate how the deformation affects the effective potential and study the marginally bound orbits (MBOs) and the marginally stable circular orbits (MSCOs).
Then, in section \ref{sec4} the rational number $q$ of a periodic orbit, related to the triplet, is calculated and the impact of deformation parameter on the energy and orbital angular momentum is explored. 
After this, a catalog of periodic orbits is concluded. 
In section \ref{sec5} we present the calculations of gravitational waveforms and comparison with Schwarzschild spacetime scenario, illustrating the influence of modification in spacetime. 
Finally, the conclusions and discussions are given in section \ref{sec6}.

\section{The non-rotating DSK black hole spacetime} \label{sec2}
Here, we briefly introduce the non-rotating DSK black hole spacetime.
The metric of the deformed spacetime reads
\begin{equation} \label{e1}
    \mathrm{d}s^{2}=-
    \left(1-\frac{2M}{r}+\alpha
    \frac{M^{3}}{r^{3}}\right)\mathrm{d}t^{2}+\frac{1+\alpha M^{3}/r^3}{1-2M/r}\mathrm{d}r^2+r^{2}(\mathrm{d}\theta^2+\sin^{2}\theta \mathrm{d}\varphi^{2}),
\end{equation}
with a parameter $\alpha$ describing the degree of deformation, which is the parameter $\alpha_{Q}$ in the non-Kerr spacetime\citep{PhysRevD.102.064041}.
Note that the event horizon is fixed at $r=2M$, independent of $\alpha$.
This spacetime can be taken as a form of an exact, vacuum solution of mixed scalar-$f(R)$ theories \citep{Suvorov2021}.

The intuition about the physical essence of the deformation can be illustrated by comparison with existing solutions. 
The $tt$-component is the limit of a particular non-vacuum black hole metric introduced by Bardeen, of which the $tt$-component reads $g_{tt}=-1+2M/r-3Mq^{2}/r^{3}+\mathcal{O}(1/r^{4})$. The Bardeen solution involves a metric coupled to a nonlinear electromagnetic field which is generated by Bardeen magnetic monopole. Therefore, the parameter $\alpha$ can be explained as a gravitational analogy of a Bardeen magnetic monopole\citep{PhysRevD.102.064041}. More detailed analysis needs tools like multipole moment expansions\citep{Thorne:1980ru}.

The equations governing the geodesics in the deformed spacetime \eqref{e1} can be derived from the Lagrangian
\begin{equation} \label{e2}
  \mathcal{L}=\frac{1}{2}g_{\mu\nu}\frac{\mathrm{d}x^{\mu}}{\mathrm{d}\tau}\frac{\mathrm{d}x^{\nu}}{\mathrm{d}\tau},
\end{equation}
where $\tau$ is some affine parameter along the geodesic.

The corresponding generalized momenta are
\begin{equation}
    p_{\mu}=\frac{\partial\mathcal{L}}{\partial\dot{x^{\mu}}}=g_{\mu\nu}\dot{x^{\nu}},
\end{equation}
or specifically,
\begin{equation}\label{e4}
    \begin{split}
        &p_{t}=-\left(1-\frac{2M}{r}+\alpha\frac{M^{3}}{r^{3}}\right)\dot{t}=-E,\\
        &p_{\varphi}=r^{2}\sin^{2}\theta\dot{\varphi}=L,\\
        &p_r=\frac{1+\alpha M^{3}/r^3}{1-2M/r}\dot{r},\\
        &p_{\theta}=r^{2}\dot{\theta},
    \end{split}
\end{equation}
where the dot denotes differentiation with respect to $\tau$, and $E$ and $L$ represent the energy and the orbital angular momentum of the geodesic, respectively. In the following sections, we only considered the geodesics in the equatorial plane $(\theta=\pi/2,\dot{\theta}=0)$ due to the isotropy of the background \eqref{e1}, and set $M=1$ for simplicity.

\section{Circular Orbits} \label{sec3}
In this section, the analysis of periodic orbits will be conducted. This analysis of circular orbits is a prerequisite task for exploring periodic orbits, since the circular orbits determine the range of existence of periodic orbits. 

\subsection{The Effective Potential}
Substituting eq. \eqref{e4} into eq. \eqref{e2}, one can get the radial equation  of motion
\begin{equation}\label{e5}
    \dot{r}^{2}+V_{\text{eff}}(r)=0,
\end{equation}
with an effective potential
\begin{equation}\label{Veff}
    V_{\text{eff}}=\frac{1}{g_{rr}}
    \left(\frac{E^2}{g_{tt}}+\frac{L_{z}^2}{g_{\varphi\varphi}}-2\mathcal{L}
    \right).
\end{equation}

The null geodesics, with $\mathcal{L}=0$, depend solely on the ratio $L_z/E$, so we simply set $E=1$ in this case.
Then the effective potential reads
\begin{equation}
    V_{\text{eff}}=
    \frac{(r-2)[L_{z}^{2}(r-2)r^{2}-r^{5}+L_{z}^{2}\alpha]}
    {[(r-2)r^{2}+\alpha](r^{3}+\alpha)}.
\end{equation}

For the time-like geodesics,  $2\mathcal{L}=-1$, the effective potential is
\begin{equation}
    V_{\text{eff}}=-\frac{(r-2)(2L_{z}^{2}r^{2}-L_{z}^{2}r^{3}+2r^{4}-r^{5}+E^{2}r^{5}-L_{z}^{2}\alpha-r^{2}\alpha)}{[(r-2)r^{2}+\alpha](r^{3}+\alpha)}.
\end{equation}

\subsection{Different Circular Orbits}
Circular orbits are solutions to 
\begin{equation} \label{circular}
    V_{\text{eff}}(r)=0, V_{\text{eff}}^\prime(r) =0, 
\end{equation}
which are rather complicated to solve directly for the effective potential \eqref{Veff}.
However, we can simplify these equations by rewriting the effective potential as $V_{\text{eff}}(r)=W(r;\alpha)V_{\text{simp}}(r)$ with
\begin{equation}
   W(r;\alpha) = \frac{r-2}
    {[(r-2)r^{2}+\alpha](r^{3}+\alpha)}.
\end{equation}
We checked that $W(r;\alpha)\neq0$ for the circular orbits we will discuss.
Thus, solutions to \eqref{circular} are equivalent to
\begin{equation}\label{simp}
    V_{\text{simp}}(r)=0,\,
    V_{\text{simp}}^\prime(r)=0.
\end{equation} 

\begin{figure*}
    \centering
    \includegraphics[width=0.5\linewidth]{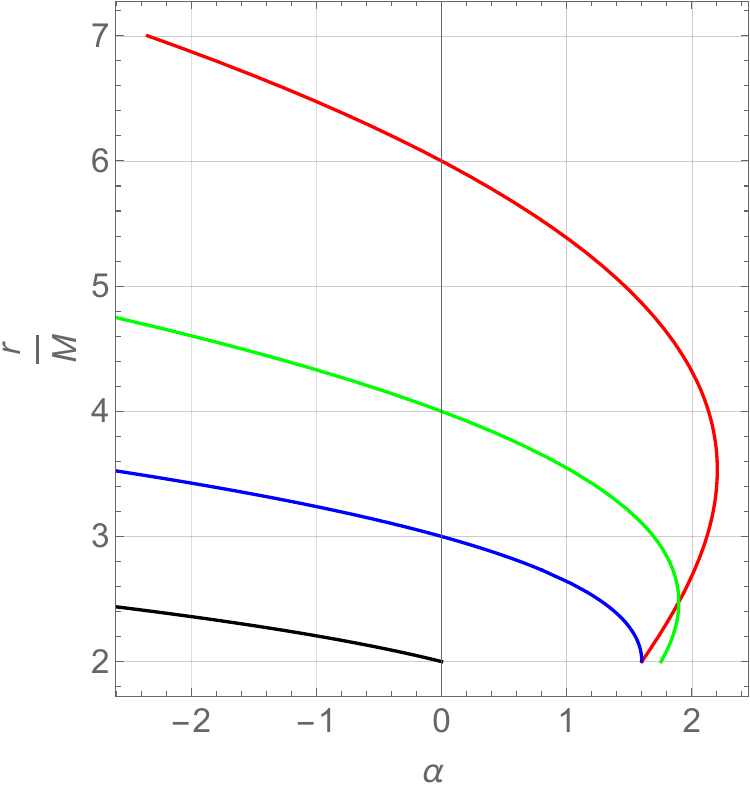}
    \caption{Different circular orbits with changes in deformation parameter $\alpha$.
    The red line, the blue line, the green line, and the black line represent the radius of MSCOs, photon orbits, MBOs and the surface of infinite redshift, respectively.
    Note that the lines inside the horizon are not shown in the figure.
    The Schwarzschild case corresponds to $\alpha=0$. 
    }
    \label{f1}
\end{figure*}

Solving this simplified equations for the null geodesic, we can obtain the so-called photon ring

\begin{equation}
    \begin{split}
        r_{\text{ph}}(\alpha)=1&+\frac{1+i\sqrt{3}}{2^{\frac{1}{3}}(-4+5\alpha+\sqrt{5}\sqrt{-8\alpha+5\alpha^{2}})^{\frac{1}{3}}}\\&+\frac{(1-i\sqrt{3})(-4+5\alpha+\sqrt{5}\sqrt{-8\alpha+5\alpha^{2}})^{\frac{1}{3}}}{2^{\frac{5}{3}}},
    \end{split}
\end{equation}
as shown in Fig. \ref{f1}. As the same as the photon ring in Schwarzschild spacetime, this null circular geodesic is unstable. However, the photon orbit disappears when $\alpha=8/5$, distinct from Schwarzschild spacetime. 

Solving the simplified equations \eqref{simp} for the time-like geodesic gives
\begin{equation}
    \begin{split}
        &E^2=\frac{2(4r^{4}-4r^{5}+r^{6}-4r^{2}\alpha+2r^{3}\alpha+\alpha^{2})}{r^{3}(-6r^{2}+2r^{3}+5\alpha)},\\
        &L_{z}^2=-\frac{-2r^{4}+3r^{2}\alpha}{-6r^{2}+2r^{3}+5\alpha}.
    \end{split}
\end{equation}

Firstly, MBOs are defined as $E(r_\text{b},\alpha)=1$, which gives a sixth-order equation about $r_\text{b}$. However, this equation can also be taken as a quadratic equation about $\alpha$, then the relation between $\alpha$ and $r_{\text{b}}$ can be solved\footnote{We drop other unreasonable roots here.}
\begin{equation}
    \alpha(r_{\text{b}})=\frac{1}{4}(8r_{\text{b}}^{2}+r_{\text{b}}^{3}-r_{\text{b}}^{5/2}\sqrt{32+r_{\text{b}}}),
\end{equation}
which gives another branch of the bound orbit when
\begin{equation}
    {\alpha\gtrsim1.7537887487646788},
\end{equation} 
This is different from the Schwarzschild spacetime, as shown in Fig. \ref{f1}. 
The two MBO branches meet together when
\begin{equation}
    {\alpha=\alpha_{\text{b}}\simeq1.8955944840993961}.
\end{equation}

Secondly, MSCOs are defined by $V^{\prime\prime}_{\text{simp}}(r_\text{s})=0$, which gives
\begin{equation}
    \alpha(r_{\text{s}})=\frac{1}{30}(20r_{\text{s}}^2+3r_{\text{s}}^3-\sqrt{-320r_{\text{s}}^4+240r_{\text{s}}^5+9r_{\text{s}}^6}).
\end{equation}
Similar to the MBOs, there exists another branch of MSCOs when $\alpha>8/5$. 
The two branches meet together when
\begin{equation}
    \alpha_\text{s}\simeq2.2033652989413897.
\end{equation}

Note that the point $(\alpha_{\text{b}},r_{\text{b}})$ where two branches of MBOs meet falls exactly on the second branch of MSCOs.
This can be explained mathematically. Where the two MBO branches (as two roots of $V^{\prime}_{\text{simp}}=0$) meet, the second-order derivative $V^{\prime\prime}_{\text{simp}}$ of this double root definitely equals zero.

 It should be noted that when $\alpha<0$, the $g_{tt}$ component undergoes a sign change outside the horizon, in other words, the surface of infinite redshift appears outside the horizon (see Fig.\ref{f1}). 
However, this case does not happen when $\alpha>0$, for which we mainly consider this situation in the following content.

\section{Periodic Orbits} \label{sec4}
In this section, a detailed discussion on the performance of periodic orbits will be conducted. A periodic orbit means that this trajectory can return to its initial state within a limited time. In the determination of the trajectories, this kind of orbits is very crucial. Generic orbit can be approximated by nearby periodic orbits because any irrational number can be infinitely approximated by rational numbers. Therefore, it can help to analyze the evolution of EMRIs.

For a periodic orbit, the radial velocity is zero at the point of the perihelion and apastron. So the value of effective potential is not always zero. However, this equation is a sixth-degree equation, and there is no root-finding formula. So numerical solution method can be used to judge the approximate distribution of roots.
It is for sure that there will be a root with value equaling to 2, which is the radius of horizon. This means that only roots of which the value is larger than 2 will be considered.

For a periodic orbit, it will return exactly to initial condition within finite time, which means that the ratio between the two frequencies of motion in the directions of r and $\varphi$ is rational. This rational number can be introduced by two frequencies $\omega_{r}$ and $\omega_{\varphi}$ which are of motion in the directions of r and $\varphi$.

\begin{equation}
    q \equiv \frac{\omega_{\varphi}}{\omega_{r}}-1=w+ \frac{v}{z}.
\end{equation}

The triplet $(z, w, v)$ is used to describe the topological characteristics of the orbit. The $z$ stands for zoom, which shows the number of leaves of the  orbit. The $w$ stands for the numbers of whirls, which shows how the orbits whirl before returning to apastron. The $v$ shows that when $z > 2$ how to sequence the movement of the object between different apastrons. It is also reasonable that when $z=1$, $v$ has the only value zero, because there is only one apastron. 

It can be proven that ${\omega_{\varphi}}/{\omega_{r}}={\Delta\varphi}/{2\pi}$, and $\Delta\varphi=\oint d\varphi$ which is the angle in the equator during one period. This number should be an integer multiple of $2\pi$. In this way, q can be obtained through an integral as below:
\begin{equation}
    q=\frac{1}{\pi}\int_{r_{p}}^{r_{a}}\frac{ \dot{\varphi}}{\dot{r}}\mathrm{d}r -1.
\end{equation}
In this integral, $r_{p}$ and $r_{a}$ are the perihelion and apastron which are the roots of the equation $V_{\text{eff}}=0$. However as mentioned before, when $\alpha>8/5$ there will be another branch of stable orbit and when $\alpha\gtrsim1.7537887487646788$ there will be another branch of bound orbit. So firstly, the new branch of bound orbit, which is called inner bound orbit in this article, should be considered since this doesn't appear in Schwarzschild spacetime. Setting $\alpha=1.8$ and $L_{z}\simeq\sqrt{13.814985530588238}$, periodic orbits can be obtained with slight modification of energy as below in Fig.\ref{f2}. We adopt the method of bisection to obtain $E$, which are double-precision float numbers with error smaller than $10^{-14}$. \footnote{Although we provided all the digits of the double precision floating-point number, one or two digits in the end are not reliable.}

\begin{figure*}
    \centering
    \subfigure{
    \includegraphics[width=0.3\linewidth]{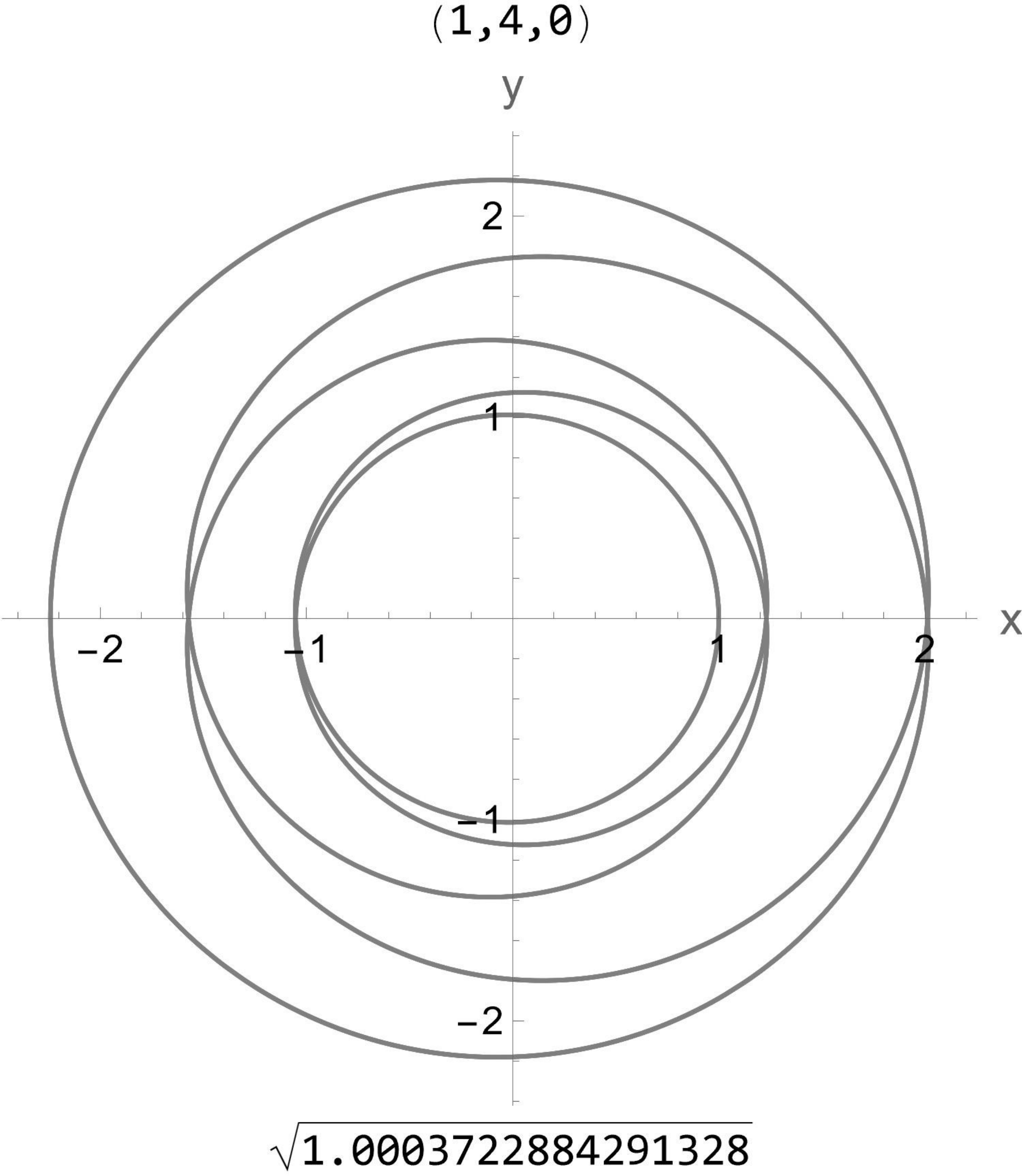}}
    \subfigure{
    \includegraphics[width=0.3\textwidth]{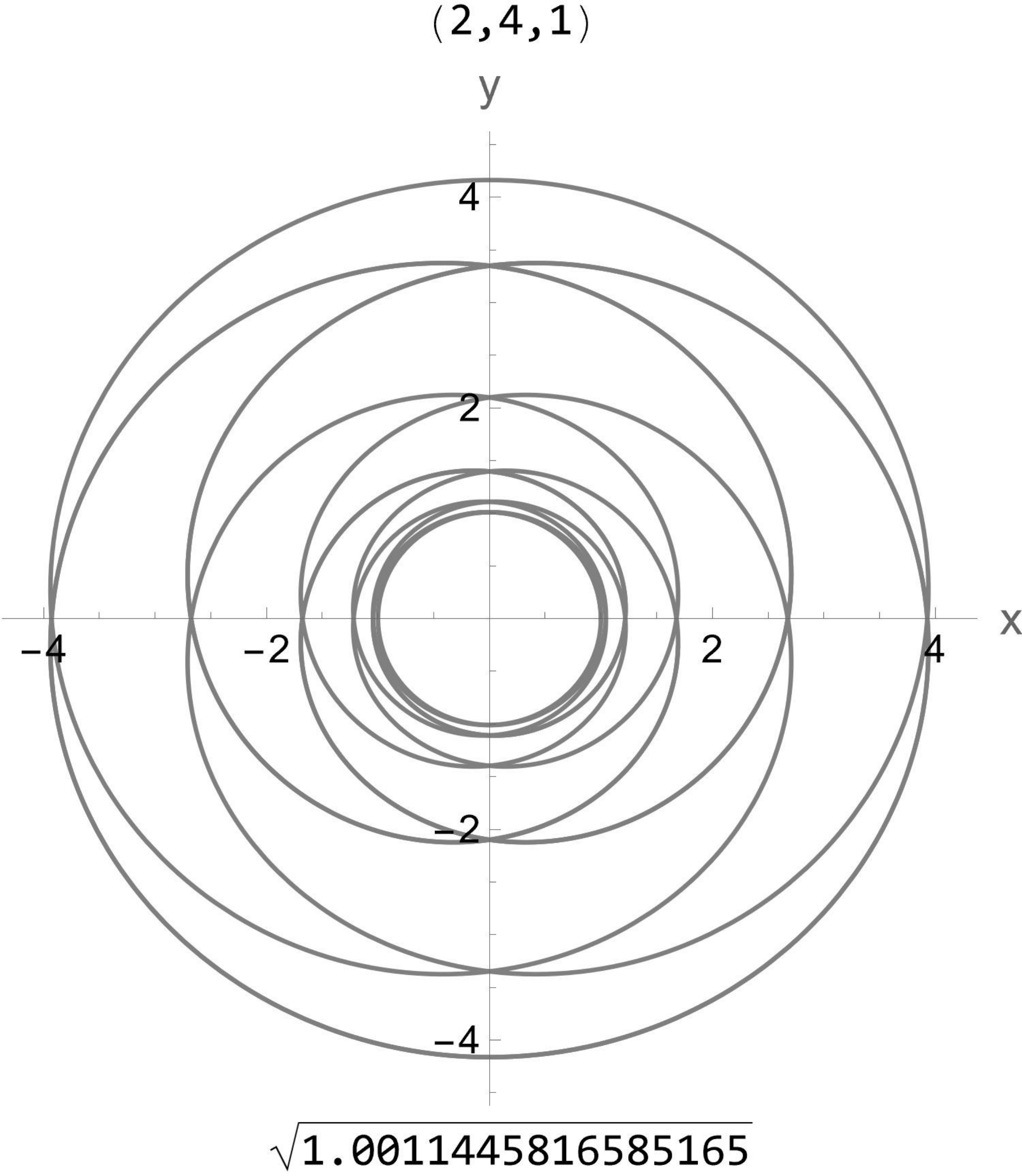}}
    \subfigure{
    \includegraphics[width=0.3\textwidth]{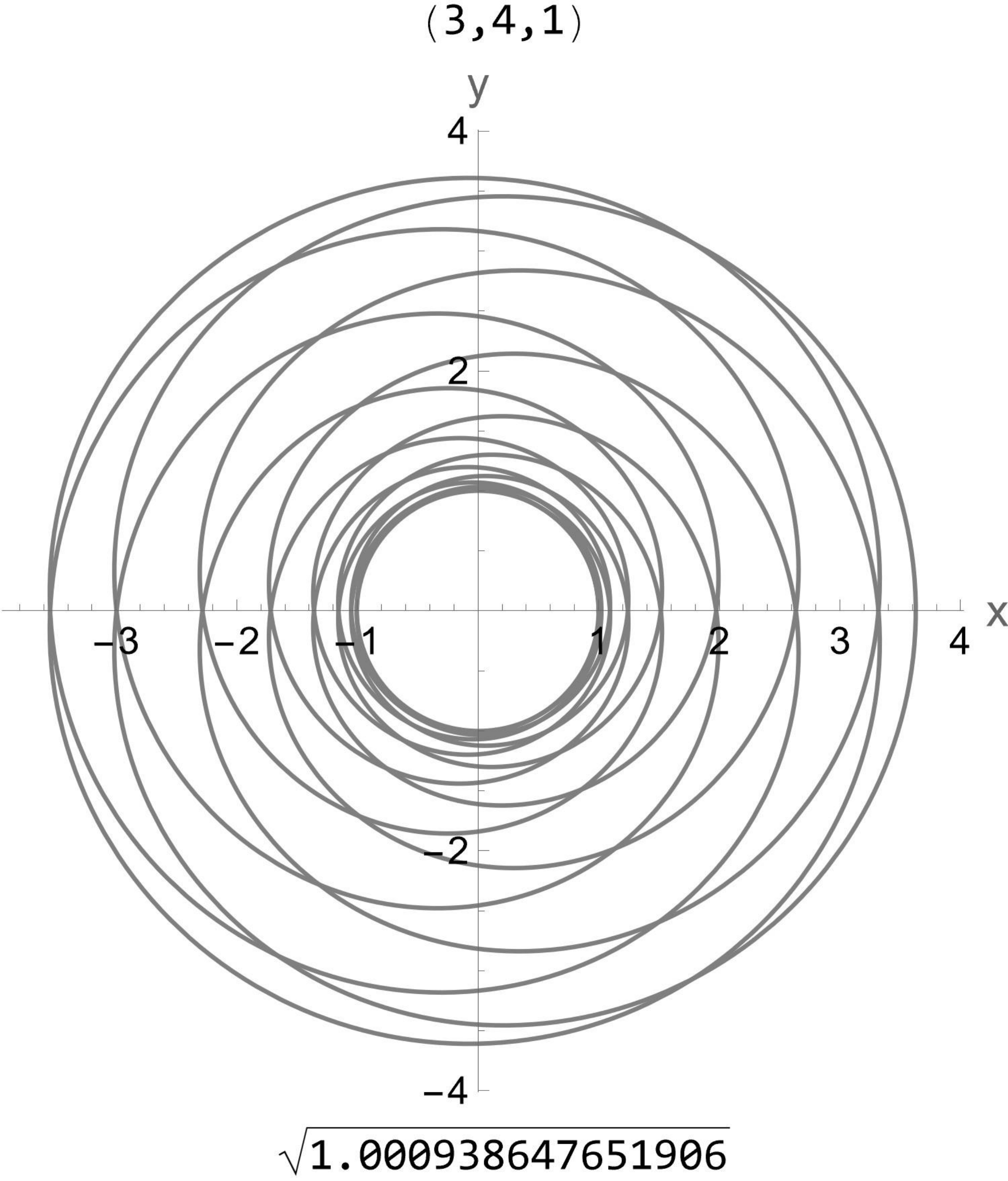}}
    \subfigure{
    \includegraphics[width=0.3\textwidth]{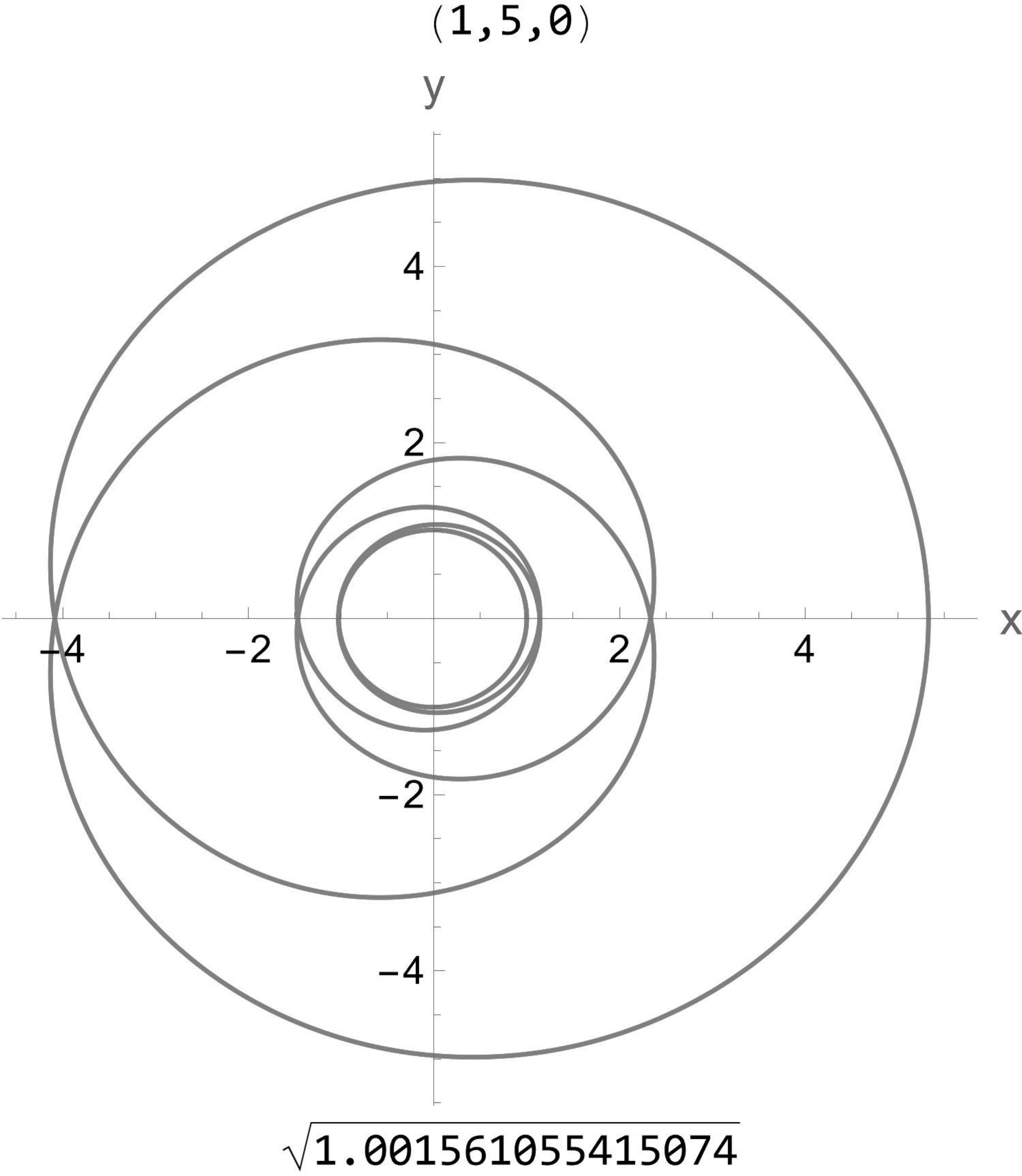}}
    \subfigure{
    \includegraphics[width=0.3\textwidth]{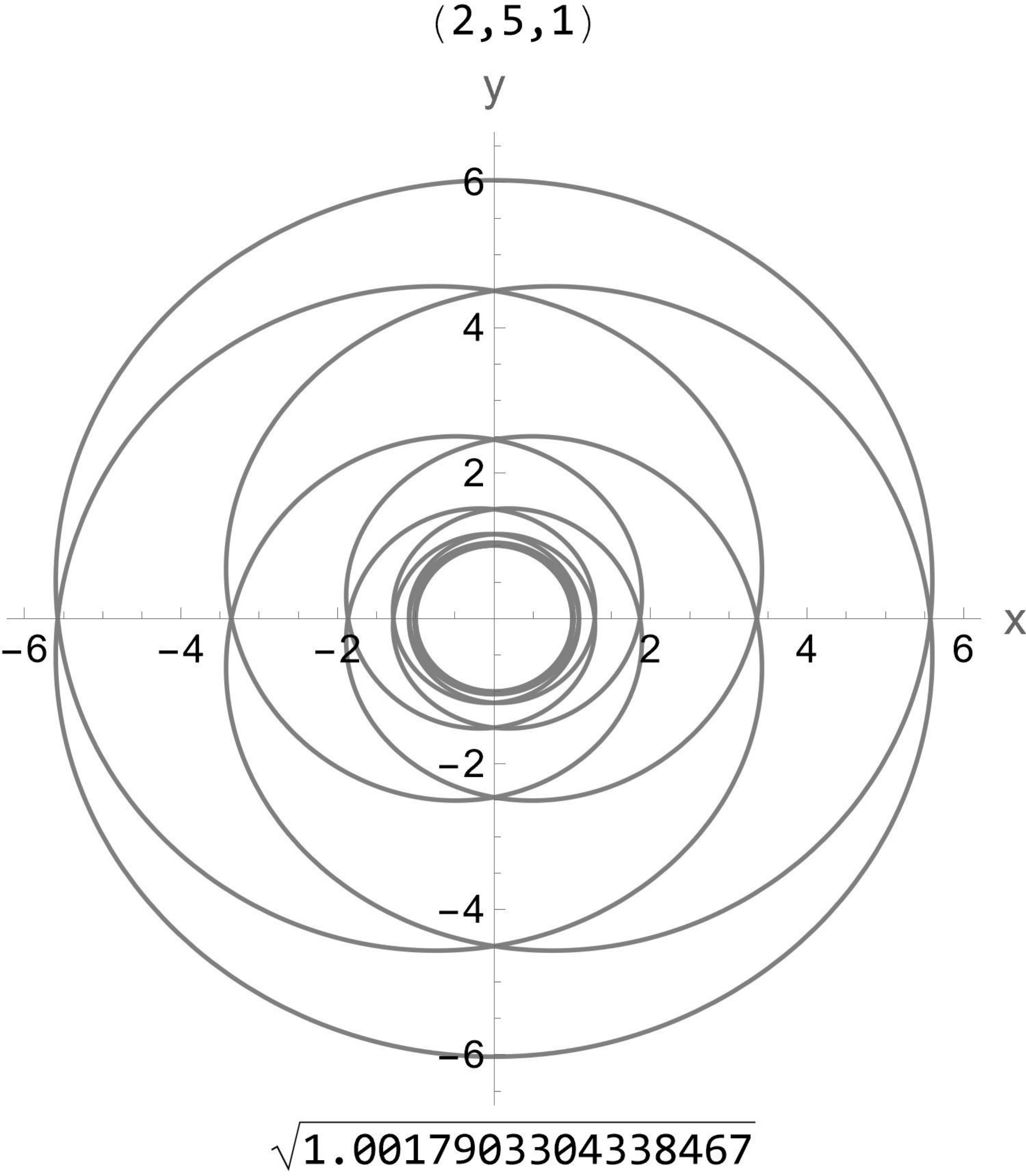}}
    \subfigure{
    \includegraphics[width=0.3\textwidth]{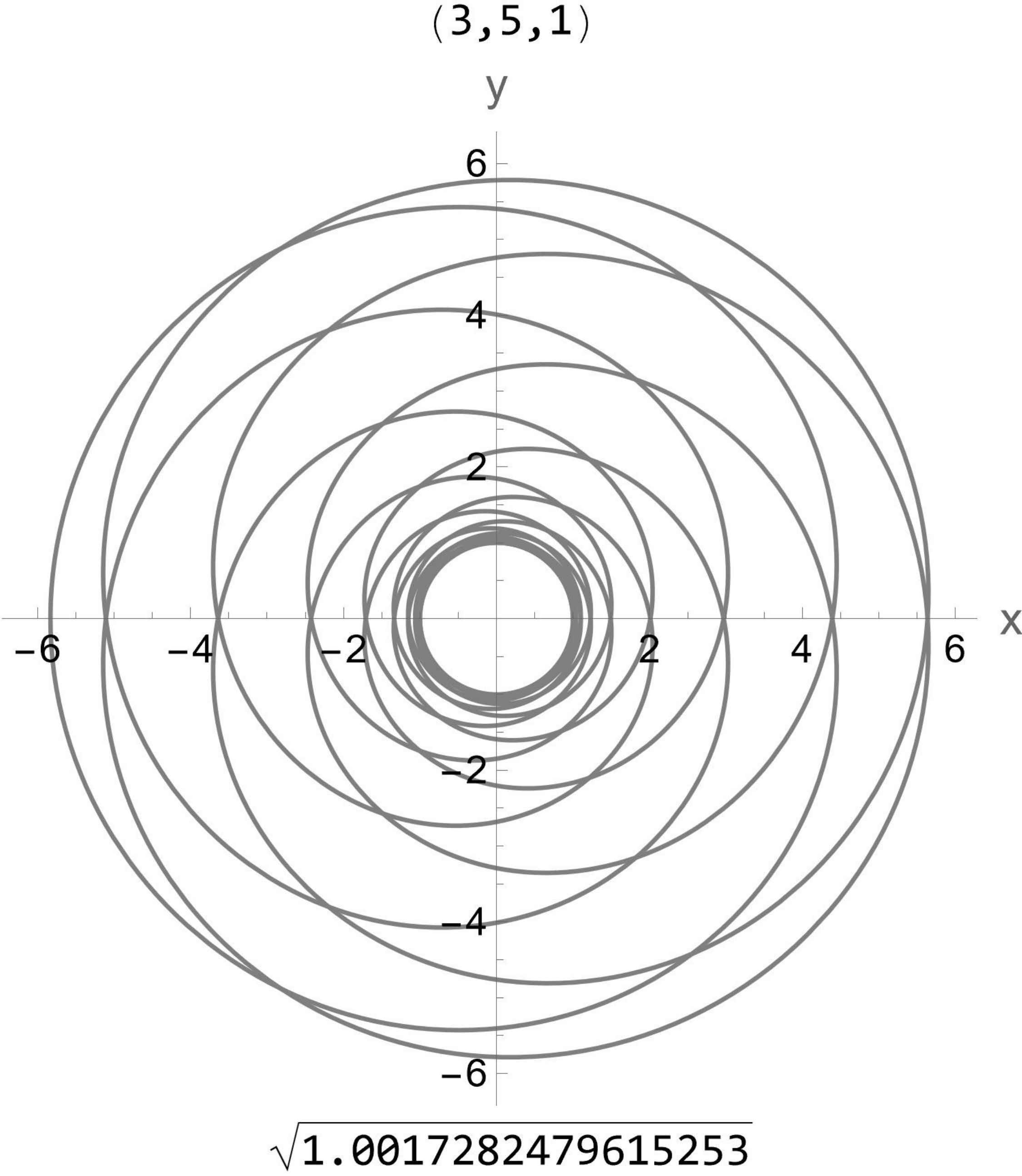}}
    \caption{Different periodic orbits are presented with deformation parameter $\alpha$ fixed at $1.8$ and the angular momentum $L_{z}$ fixed at $\sqrt{13.814985530588238}$. The triplet array and energy is shown together with corresponding orbit in every subgraph.}
    \label{f2}
\end{figure*}

Actually in this situation, the perihelion and apastron are quite close to each other. Therefore, coordinate transformation has been performed here, after which each point is the 20th power of the ratio between it and the perihelion. Another difference is that the integer $q$ seemingly cannot be smaller than 4\footnote{Perhaps due to accuracy, we cannot obtain a rational number ratio less than four in the situation we consider.}. Since orbits are equivalent to ergodic trajectories when the value of integer is large enough, the $q$ is usually set to be smaller than 6. This situation is so extreme that it may have limited practical value. 

The other branch, which is referred as outer bound orbit, is the next to be discussed. According to the effective potential, there are two regions in which the orbit can exist. One is near the horizon, which is accidentally to be the perihelion of the orbit. With slight change of the energy, the orbit will gradually enter the horizon as shown below in Fig.\ref{f3}. This means that the only orbit in this region is the bound orbit and there is no possibility for periodic orbits. So this region does not need to be considered.

\begin{figure*}
    \centering
    \subfigure[1]{
    \includegraphics[width=0.31\textwidth]
    {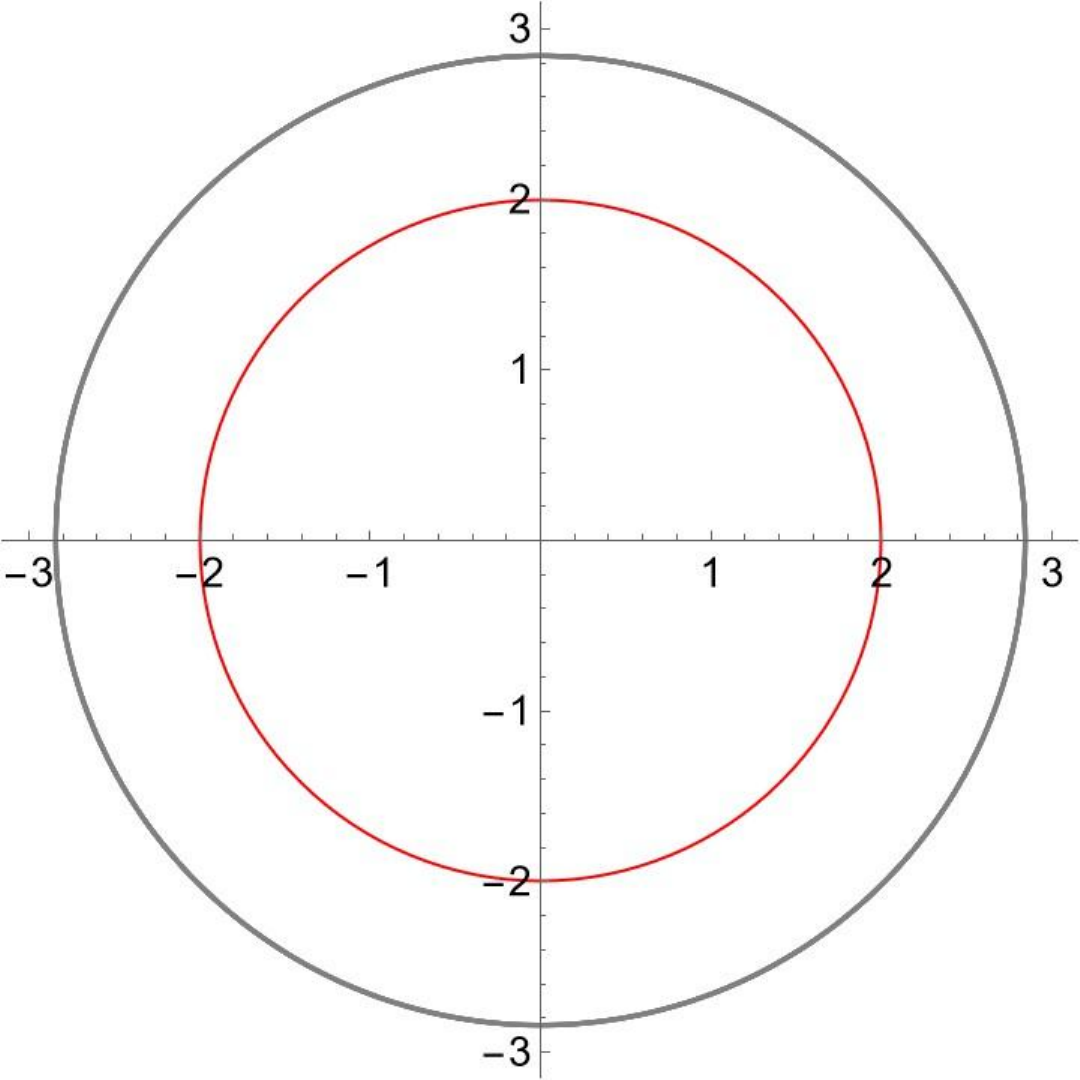}}
    \subfigure[$\sqrt{0.9999}$]{
    \includegraphics[width=0.31\textwidth]
    {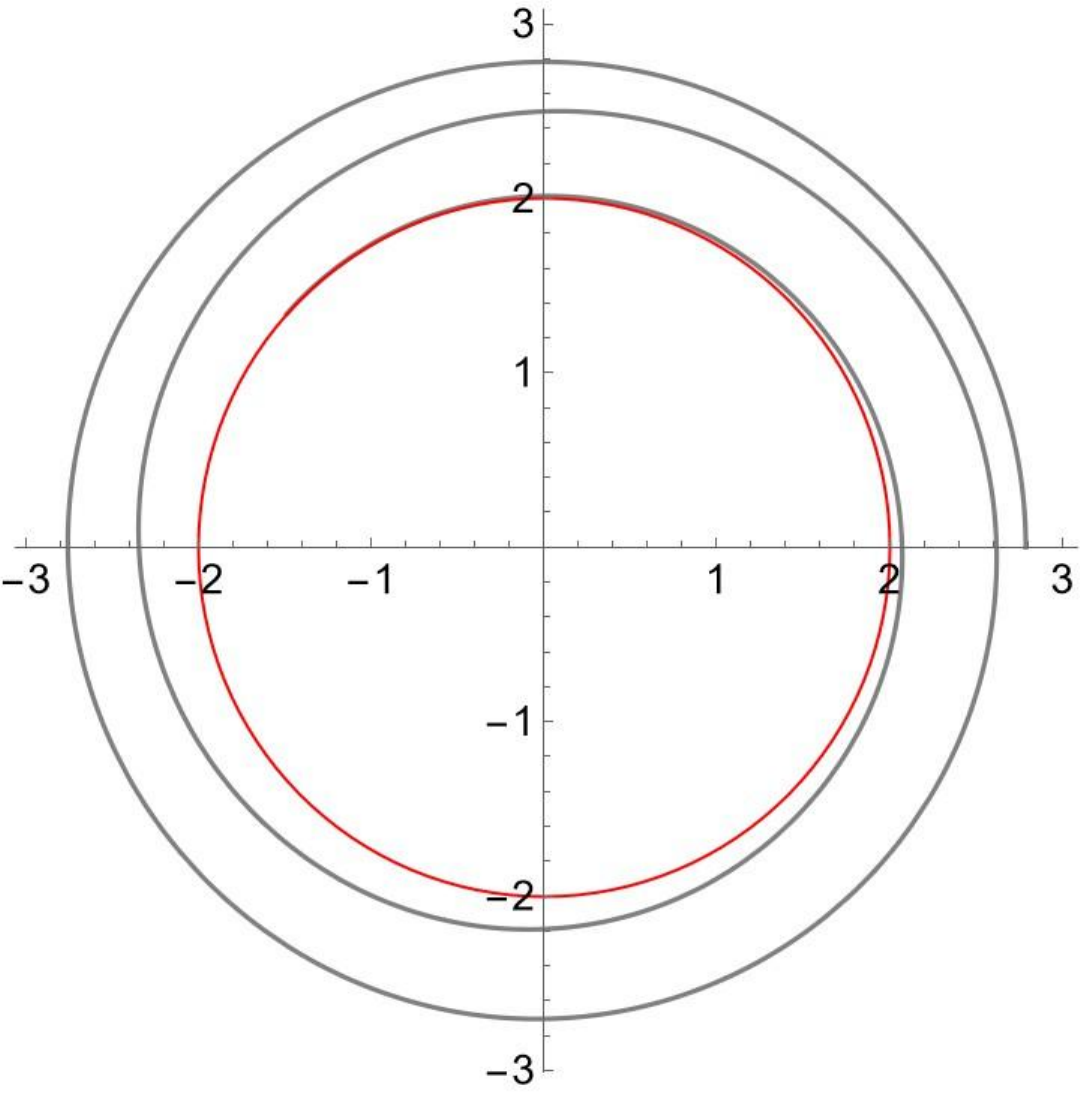}}
     \subfigure[$\sqrt{1.0001}$]{
    \includegraphics[width=0.31\textwidth]
    {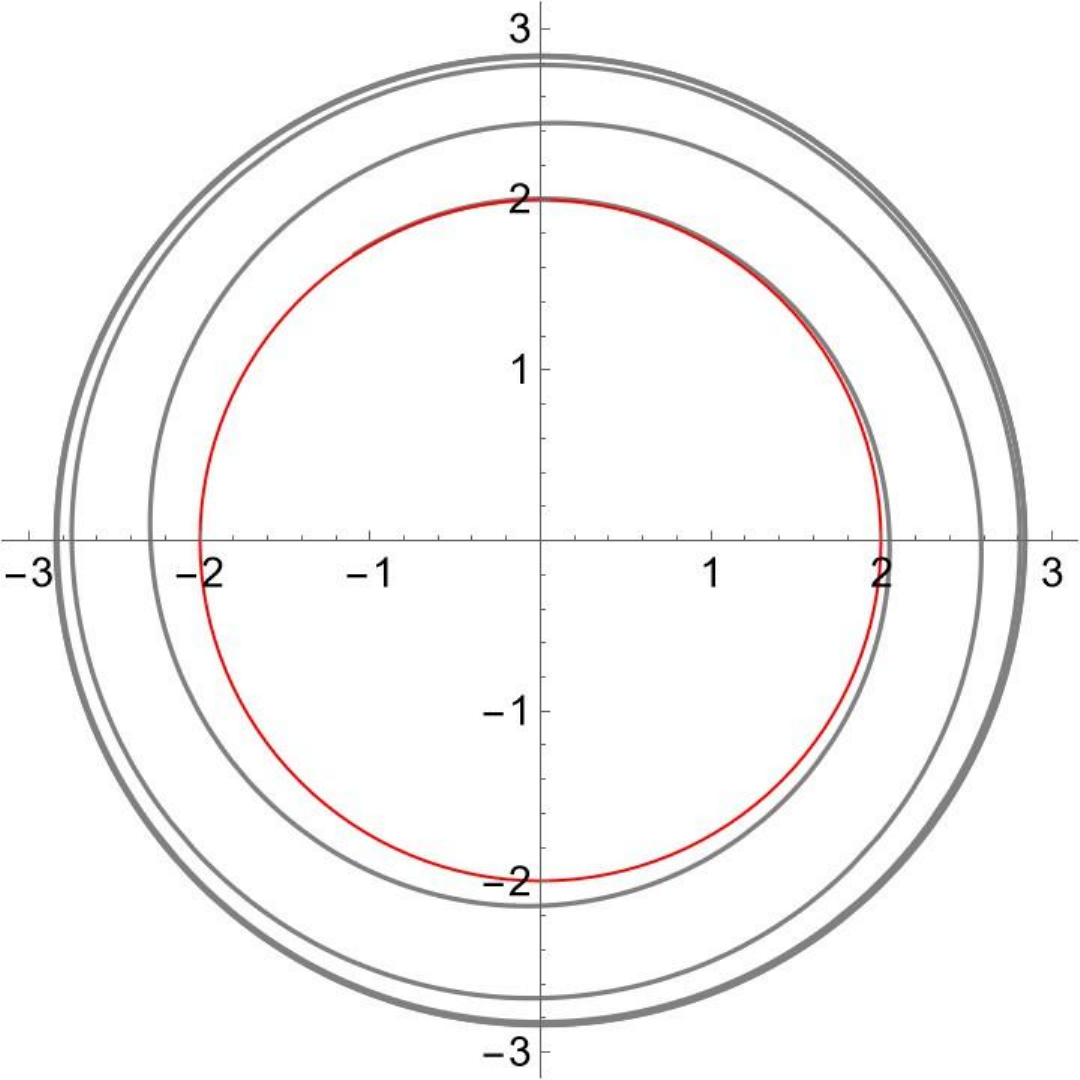}}
    \caption{Different orbits are presented with deformation parameter $\alpha$ fixed at $1.8$ and the angular momentum $L_{z}$ fixed at $\sqrt{13.814985530588238}$. The energy is shown below in every subgraph. It clearly shows the unstable behavior passing through the horizon.}
    \label{f3}
\end{figure*}

\begin{figure*}
    \centering
    \subfigure{
    \includegraphics[width=0.2\textwidth]
    {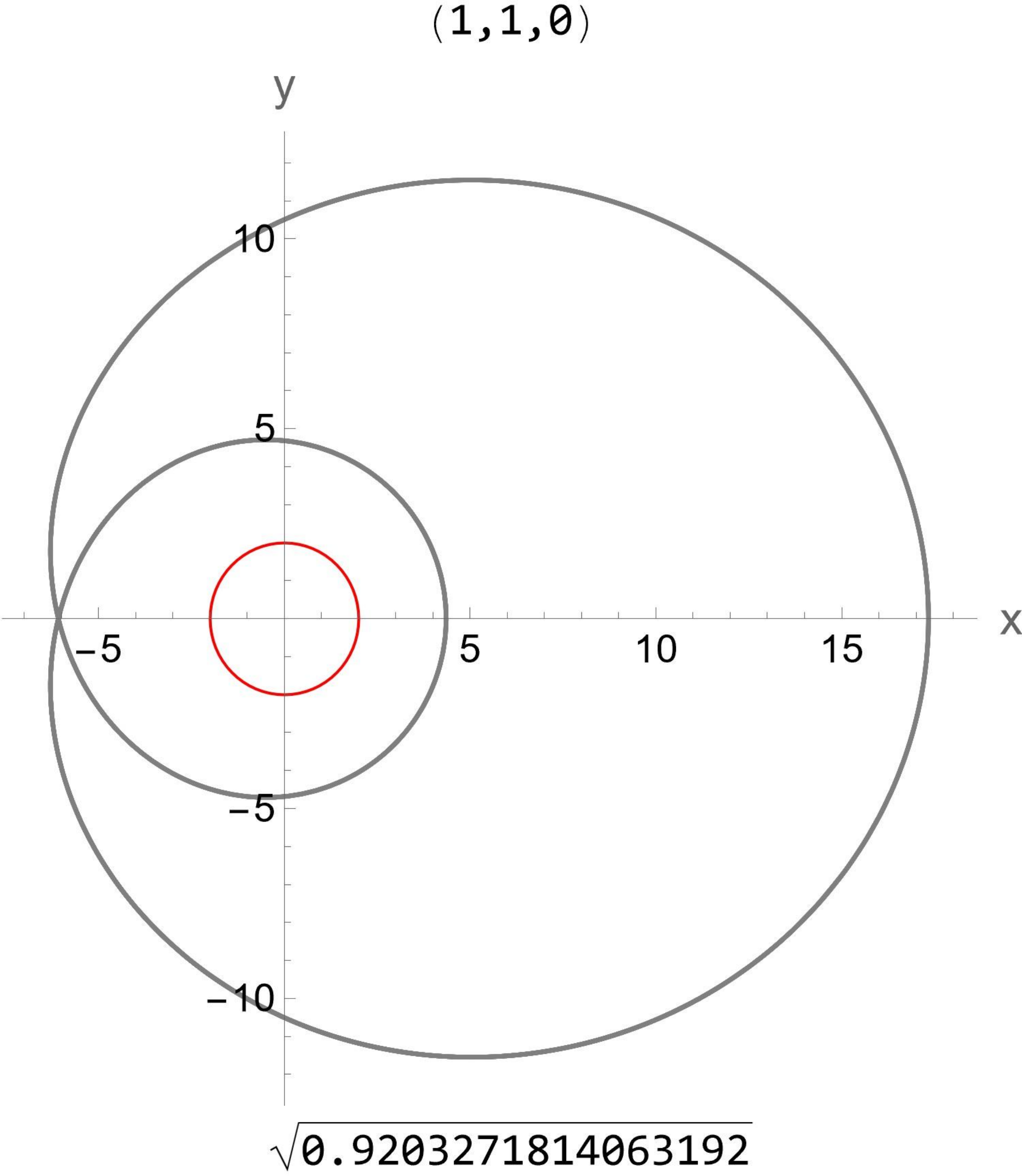}}
    \subfigure{
    \includegraphics[width=0.2\textwidth]
    {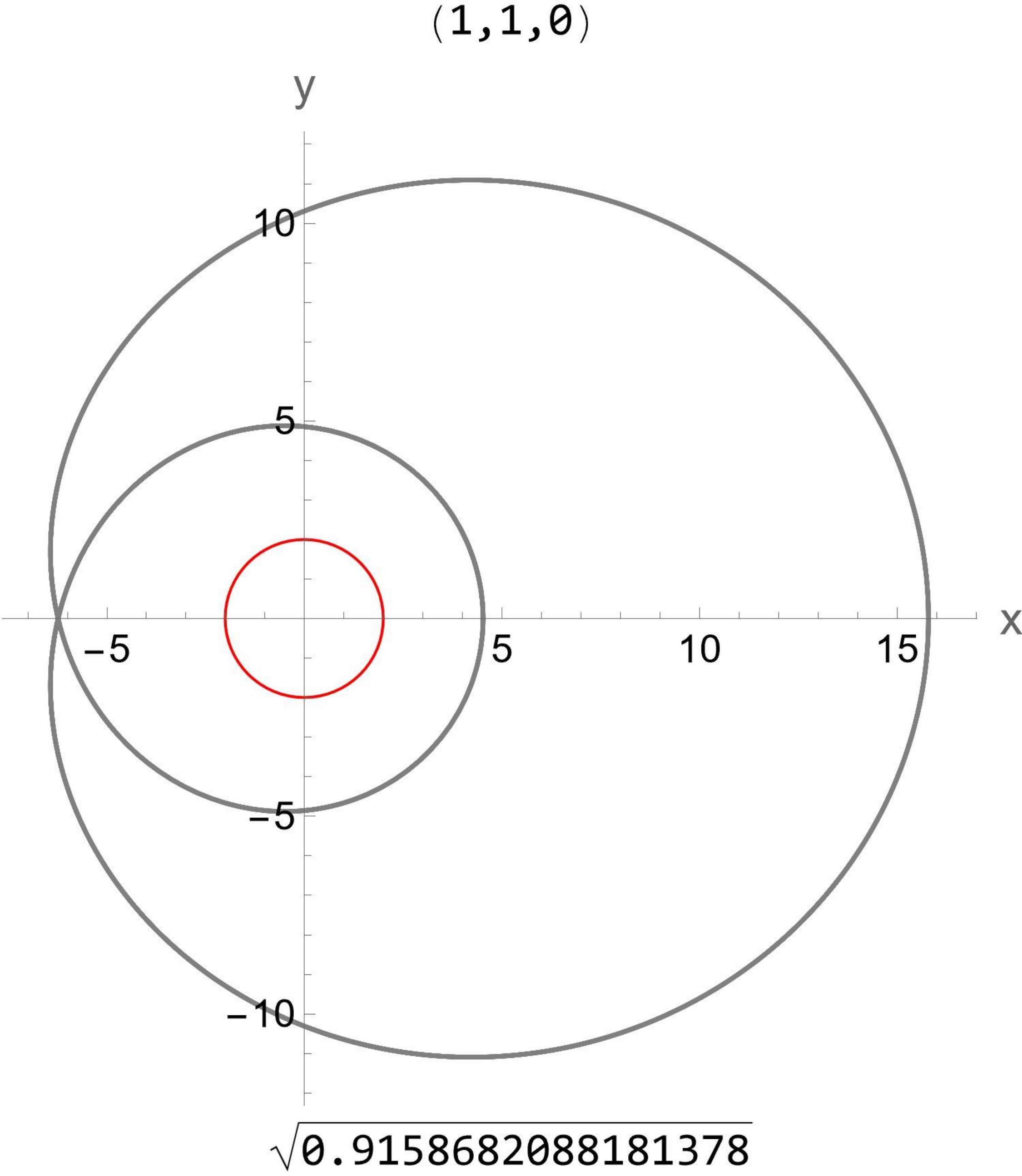}}
    \subfigure{
    \includegraphics[width=0.2\textwidth]
    {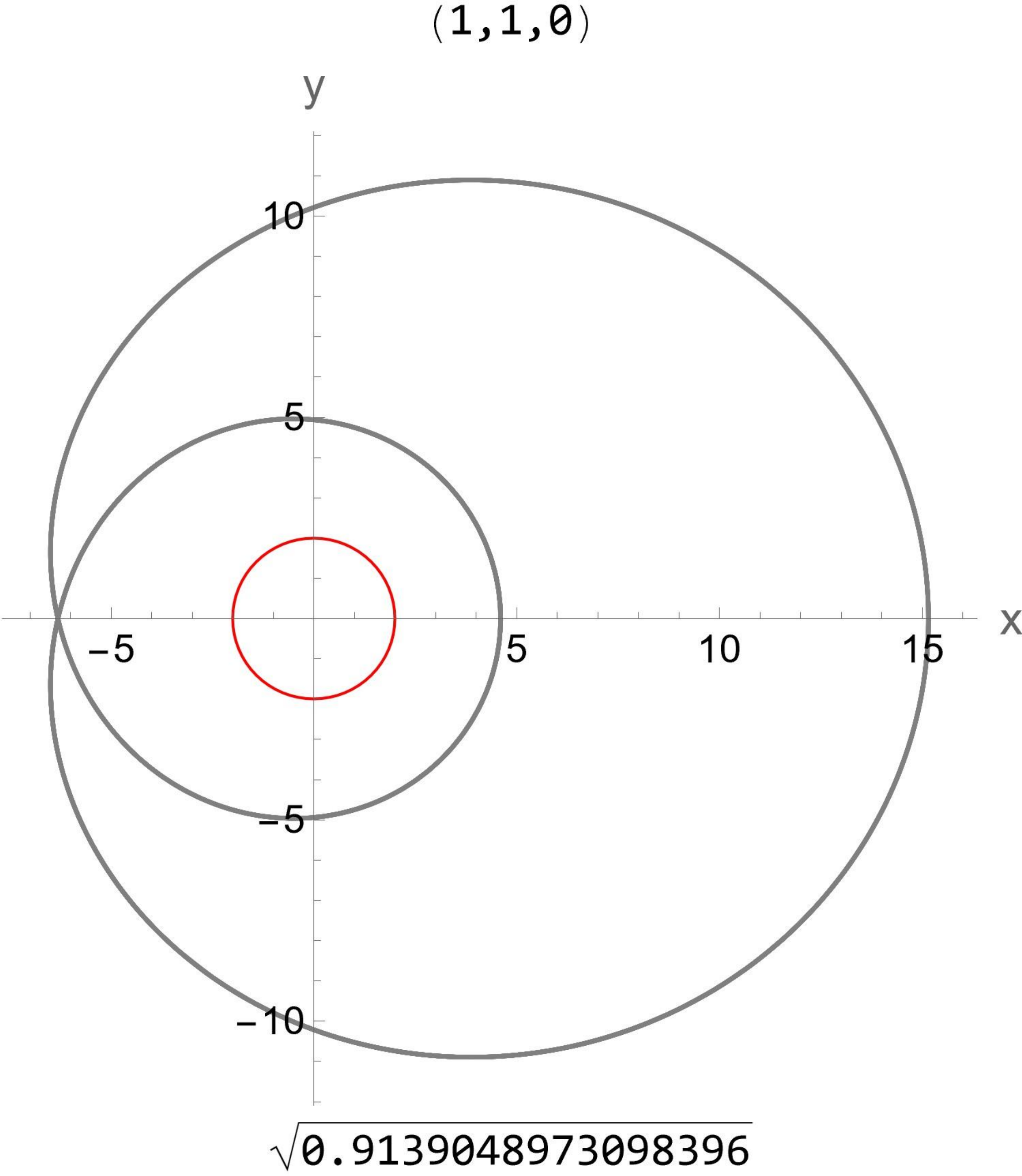}}
    
    \subfigure{
    \includegraphics[width=0.2\textwidth]
    {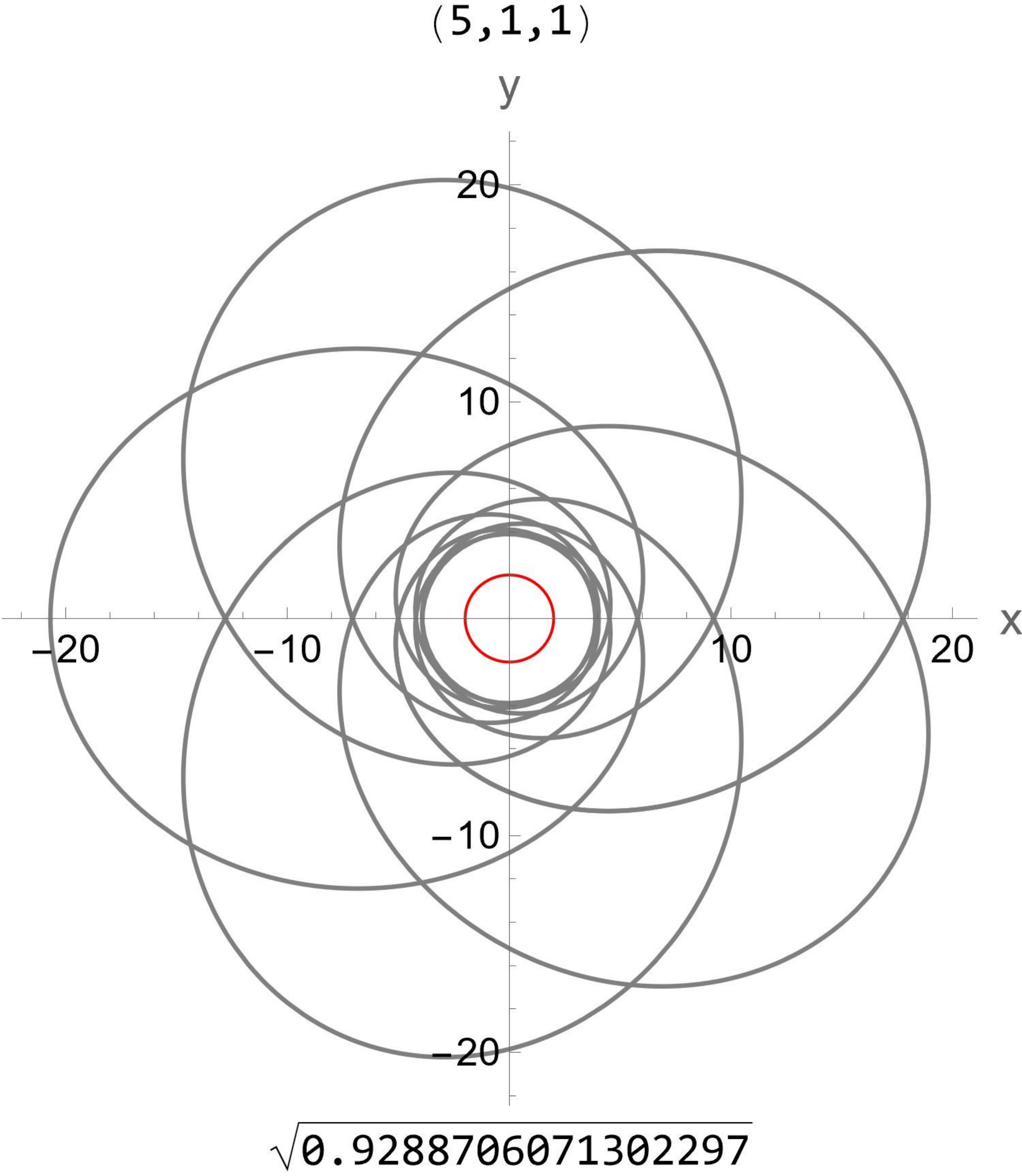}}
    \subfigure{
    \includegraphics[width=0.2\textwidth]
    {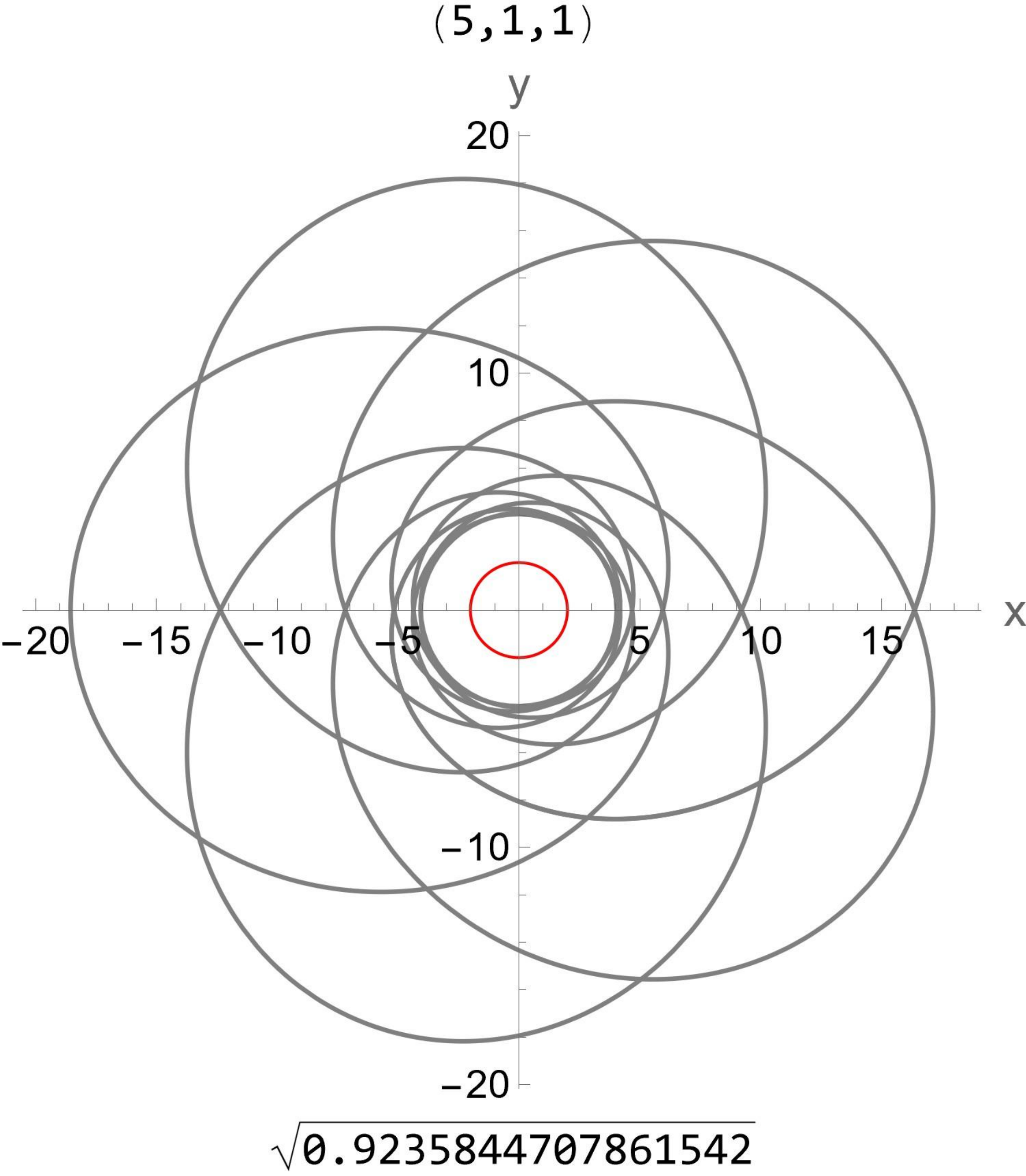}}
    \subfigure{
    \includegraphics[width=0.2\textwidth]
    {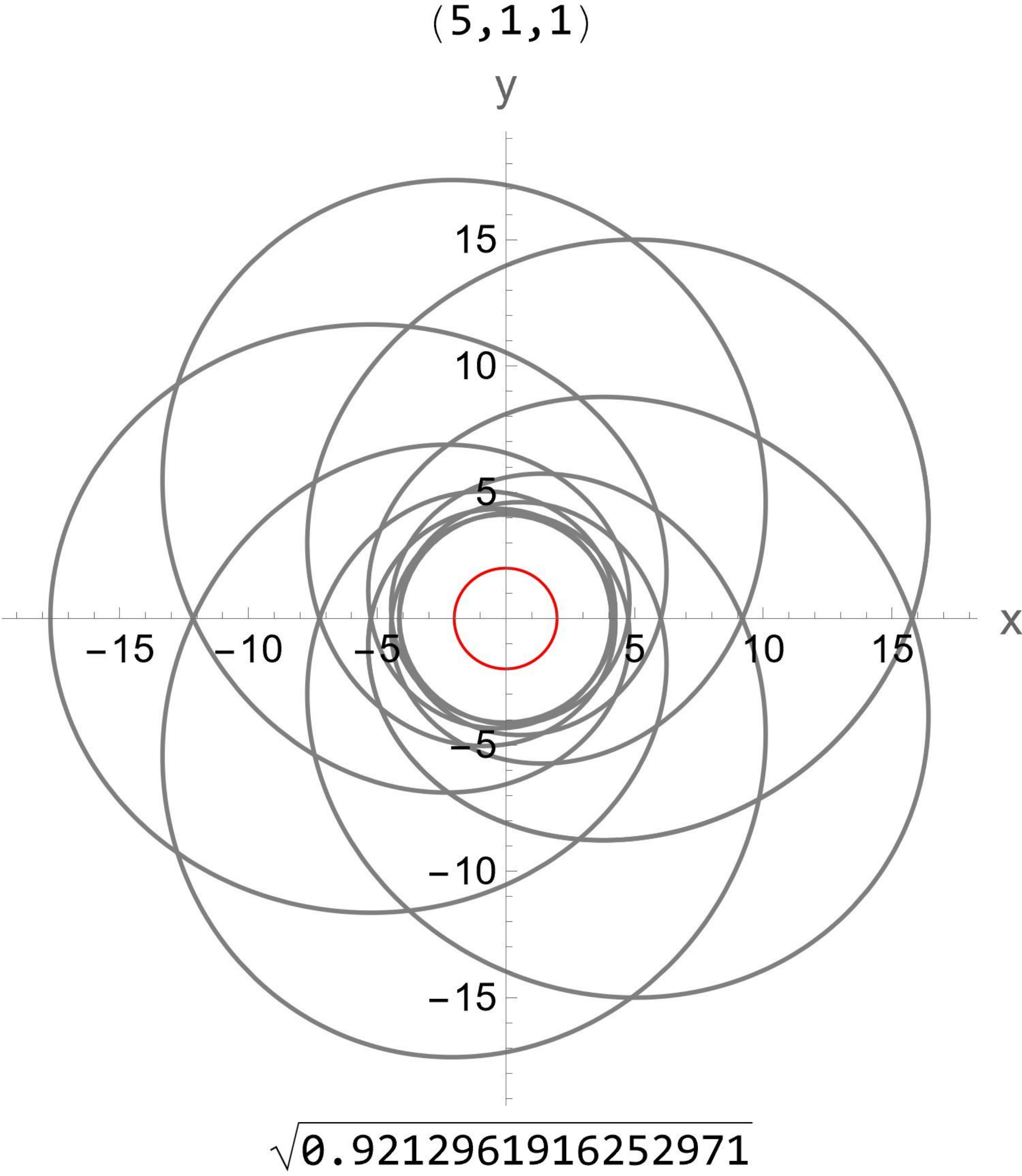}}
    
    \subfigure{
    \includegraphics[width=0.2\textwidth]
    {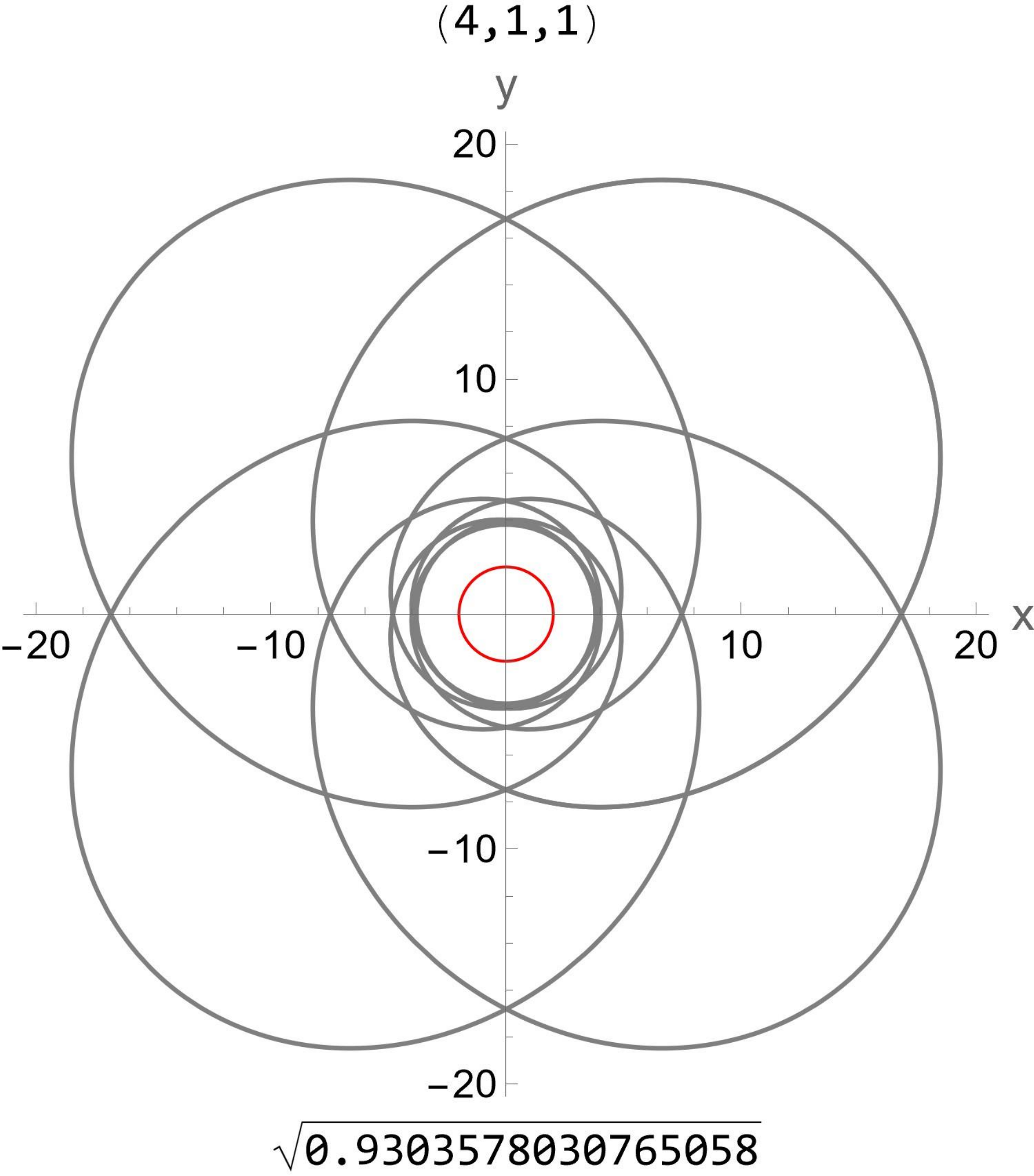}}
    \subfigure{
    \includegraphics[width=0.2\textwidth]
    {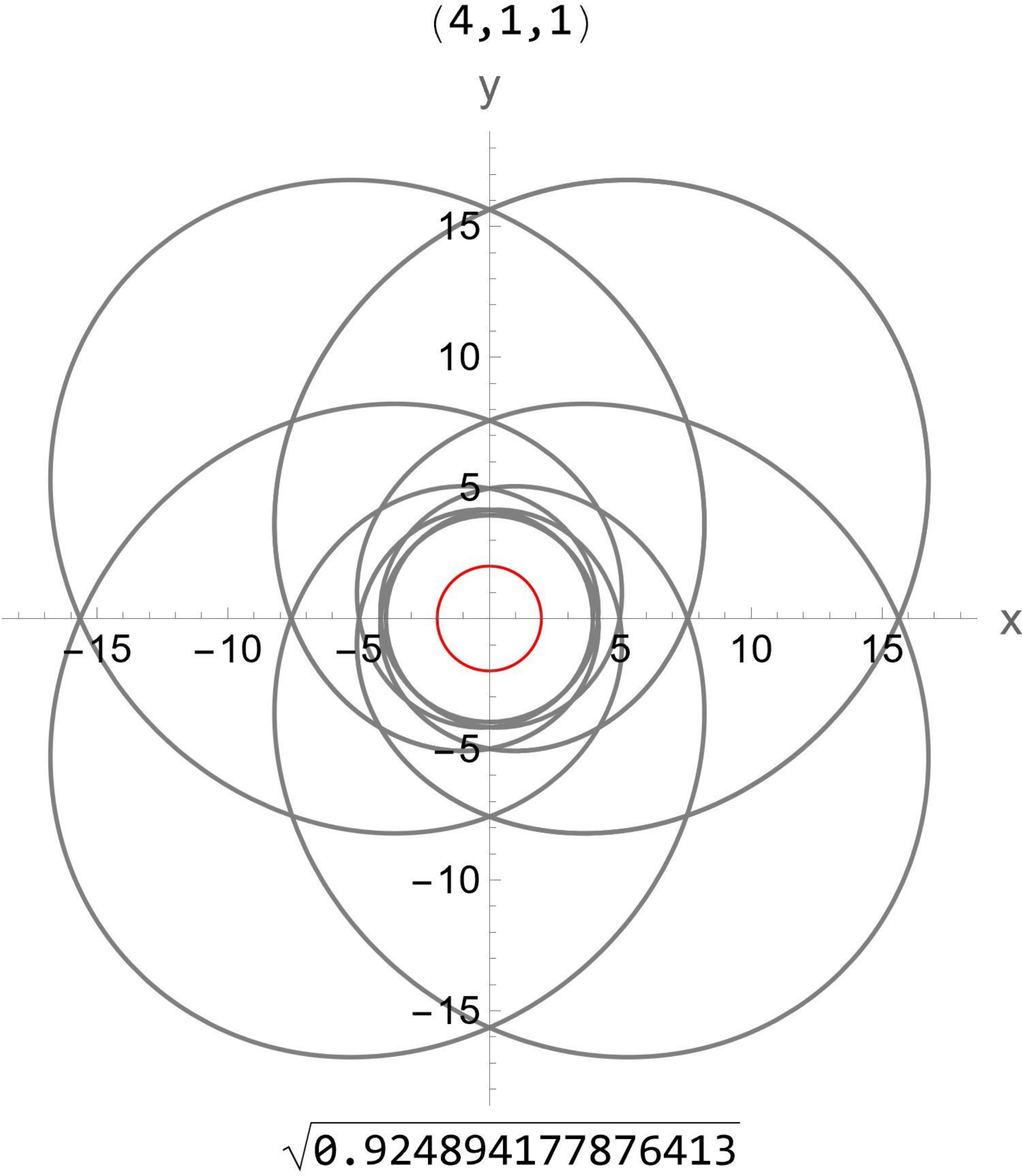}}
    \subfigure{
    \includegraphics[width=0.2\textwidth]
    {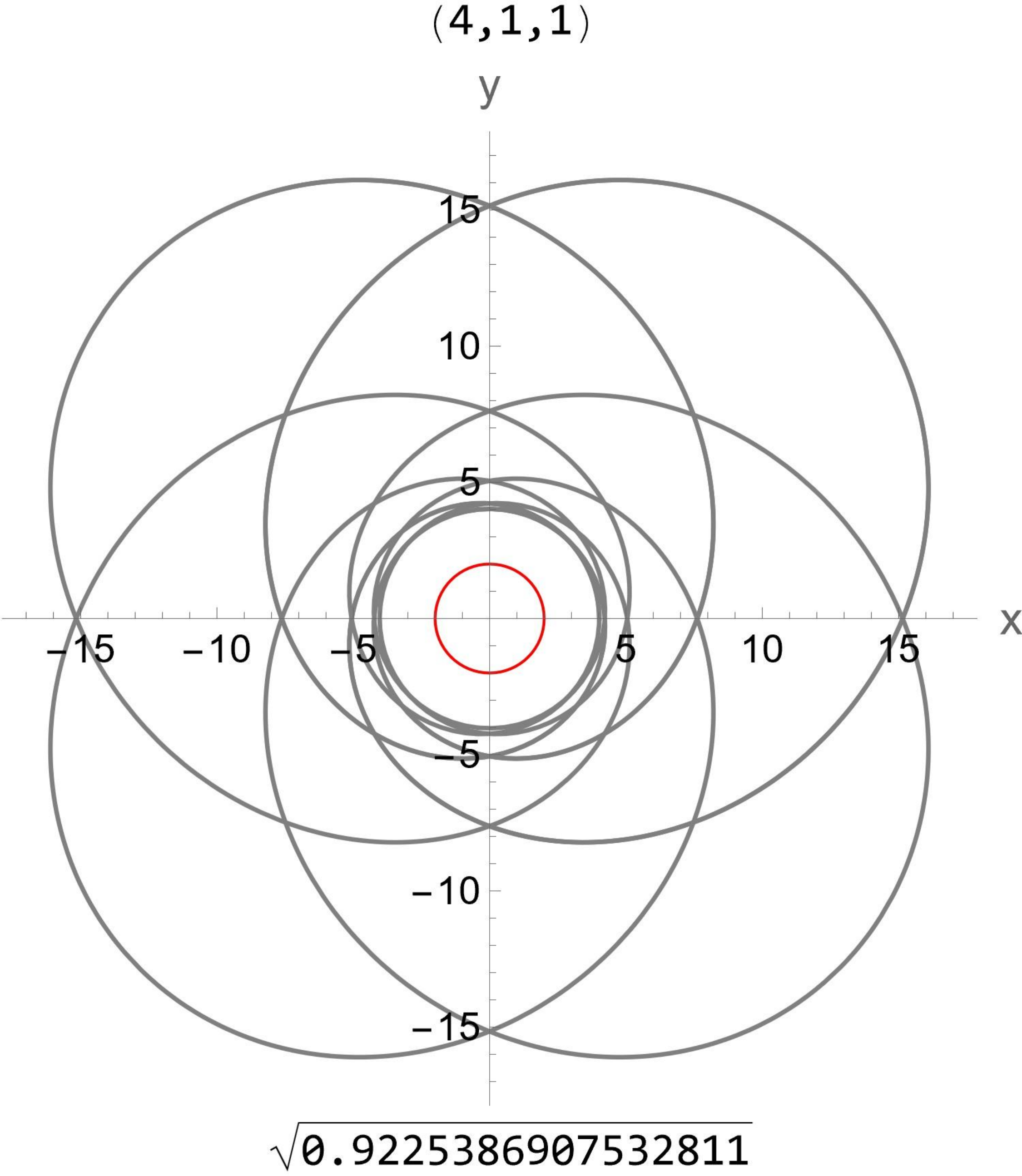}}
    
    \subfigure{
    \includegraphics[width=0.2\textwidth]
    {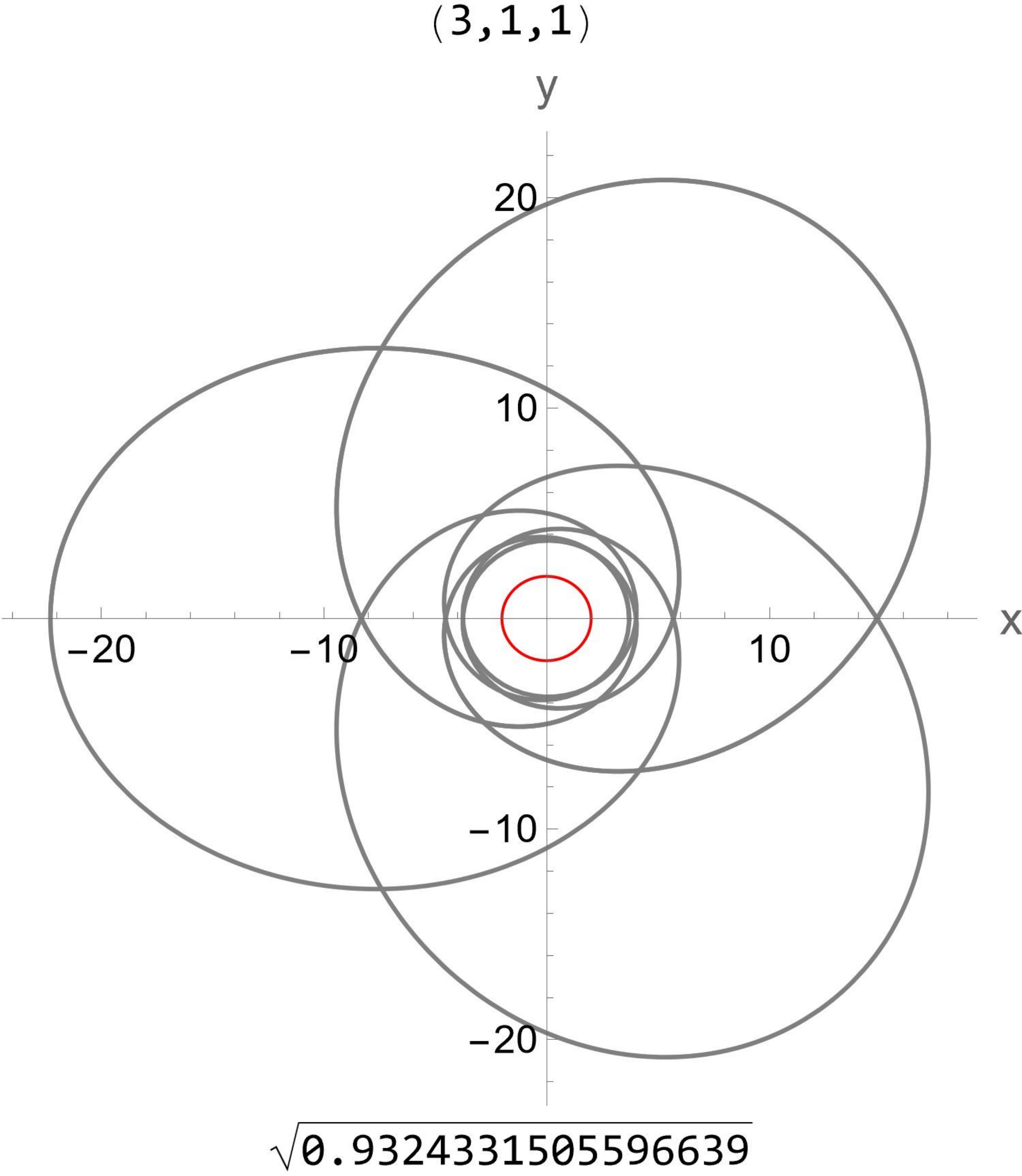}}
    \subfigure{
    \includegraphics[width=0.2\textwidth]
    {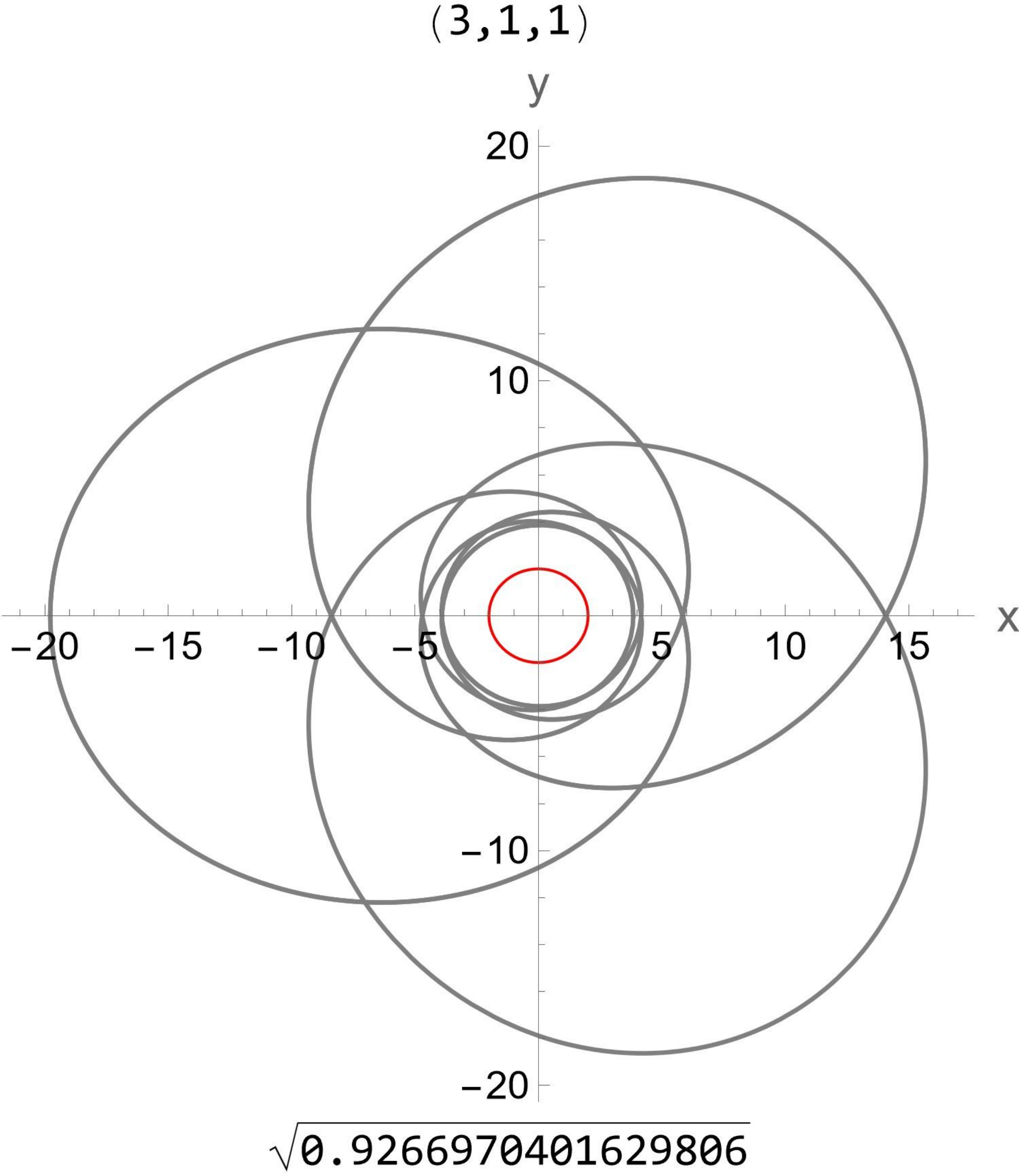}}
    \subfigure{
    \includegraphics[width=0.2\textwidth]
    {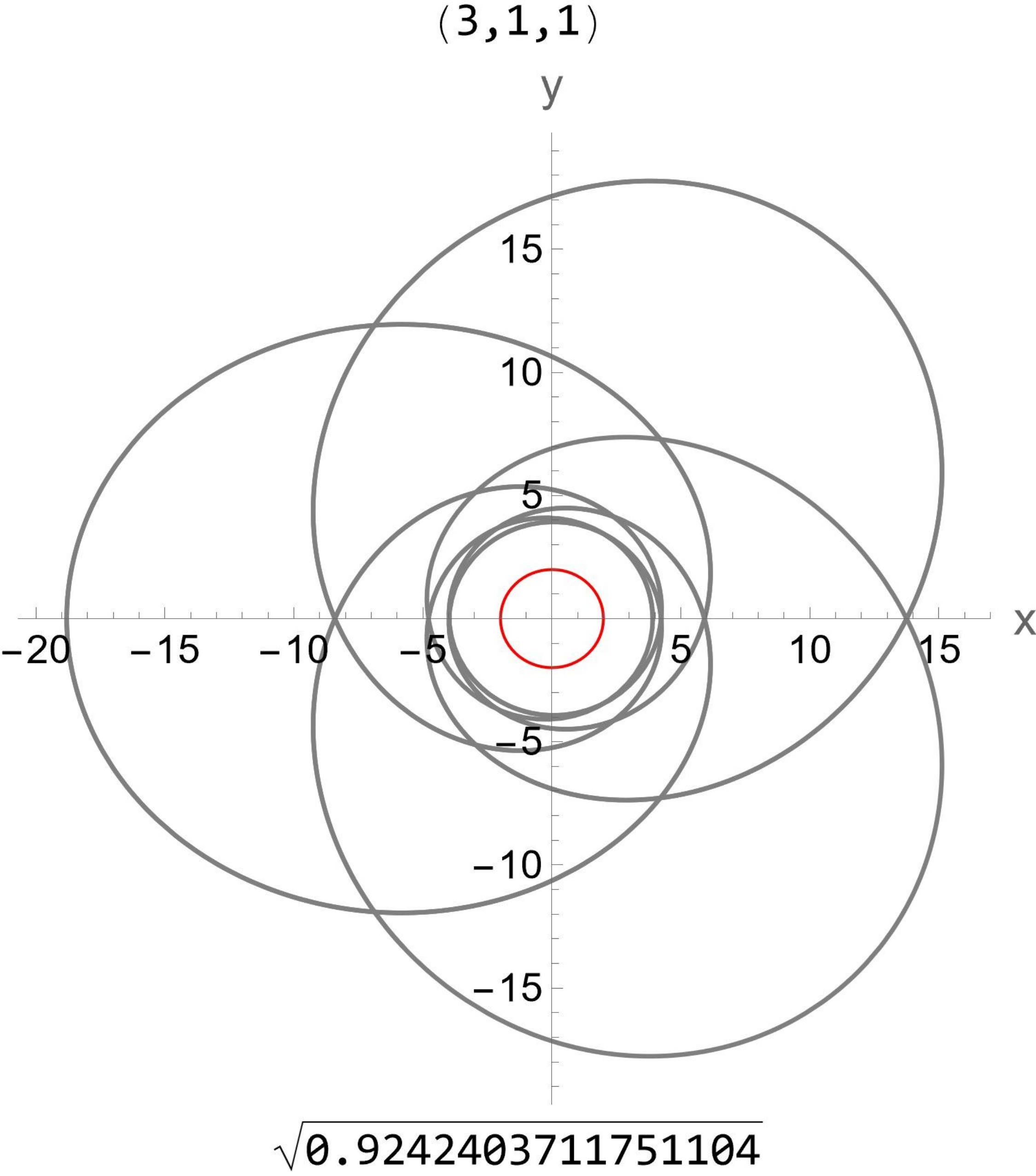}}
    
    \subfigure{
    \includegraphics[width=0.2\textwidth]
    {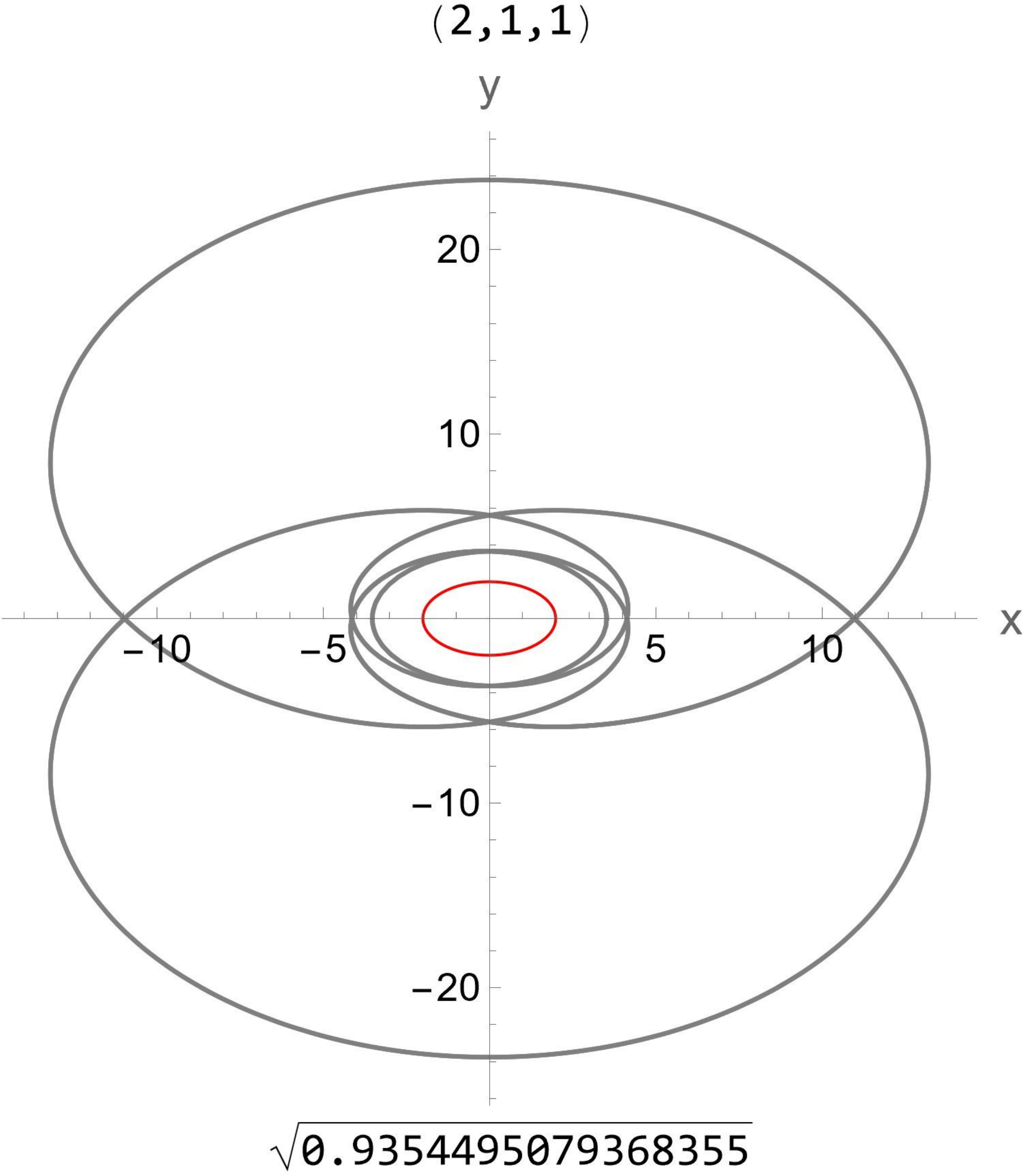}}
    \subfigure{
    \includegraphics[width=0.2\textwidth]
    {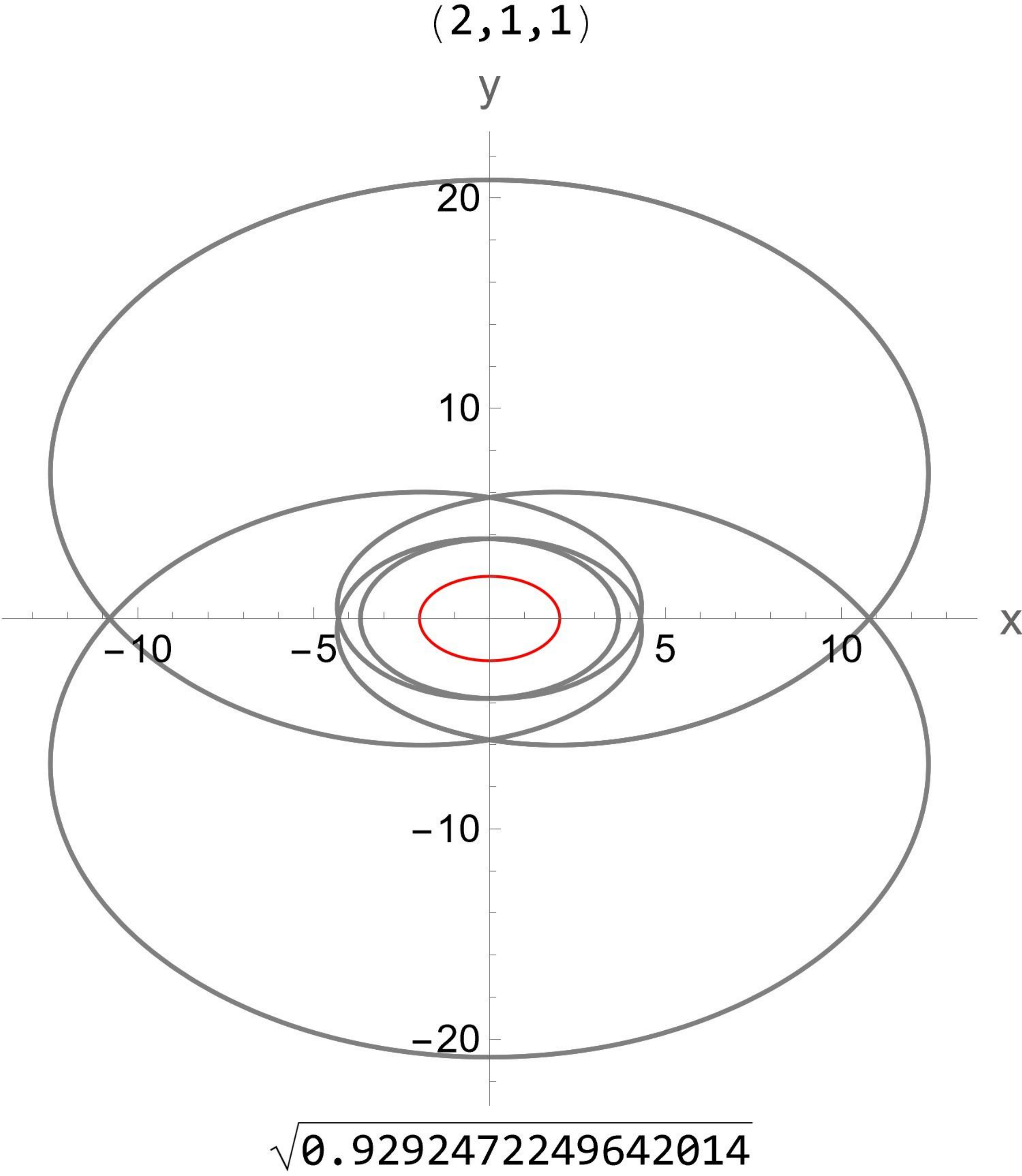}}
    \subfigure{
    \includegraphics[width=0.2\textwidth]
    {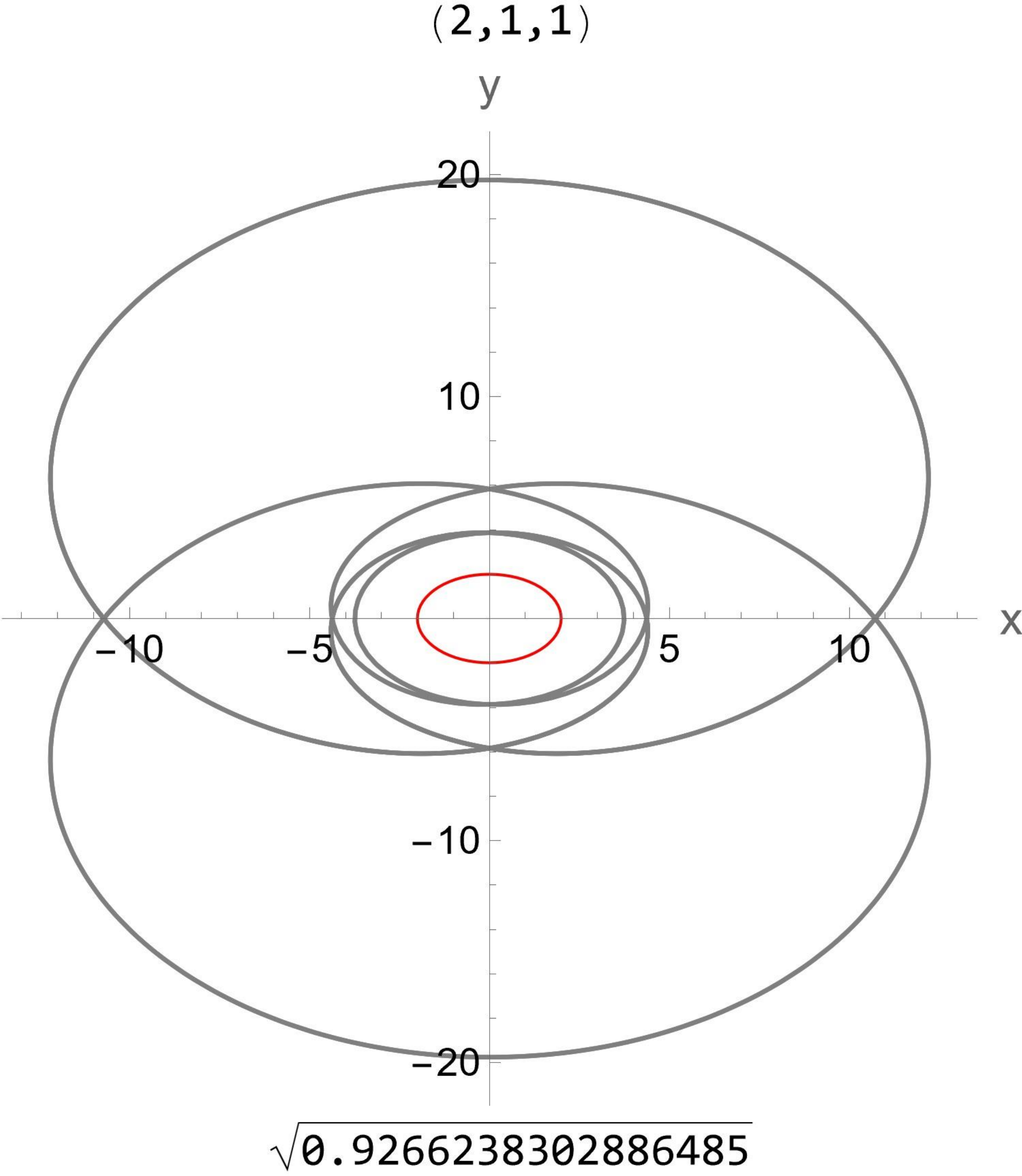}}
    \caption{Different periodic orbits are presented with the angular momentum $L_{z}$ fixed at $\sqrt{12}$. The values of deformation parameter $\alpha$ are $1.8955944840993961$, $1.8$ and $1.7537887487646788$, respectively. The values of the triplet array for each orbit are plotted also.}
    \label{f4}
\end{figure*}

In the other region, of which the perihelion is larger than 2, periodic orbits can exist. To demonstrate the influence of parameters on the orbit, the angular momentum is fixed at a constant value of $\sqrt{12}$. The integer is from 1 to 1.5 for each set of orbits. It can be found that as the value of deformation parameter increases, the perihelion decreases and the apastron increases as Fig.\ref{f4} shows.

\begin{figure*}
    \centering
    \subfigure{
    \includegraphics[width=0.25\textwidth]
     {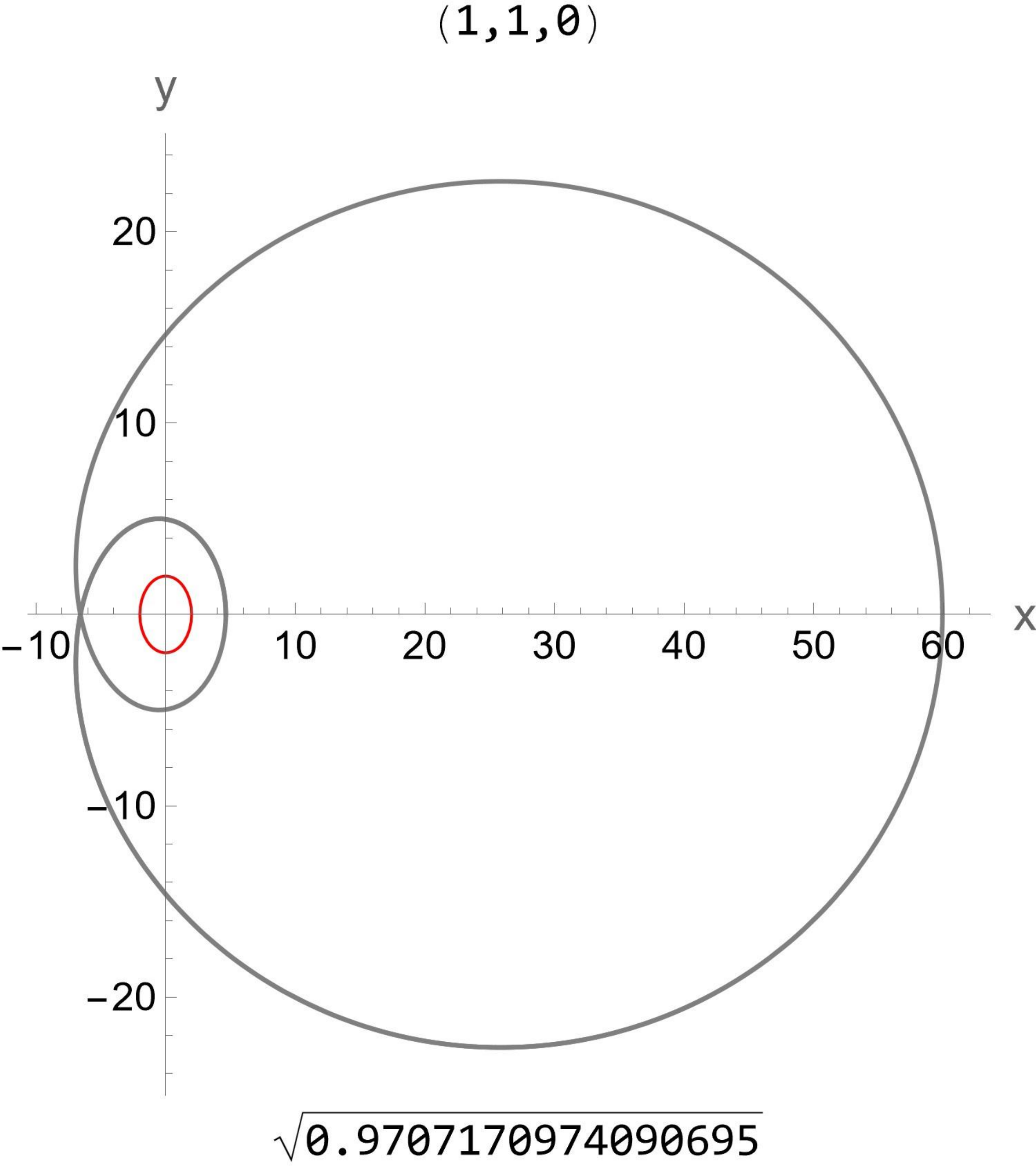}}
    \subfigure{
    \includegraphics[width=0.25\textwidth]
     {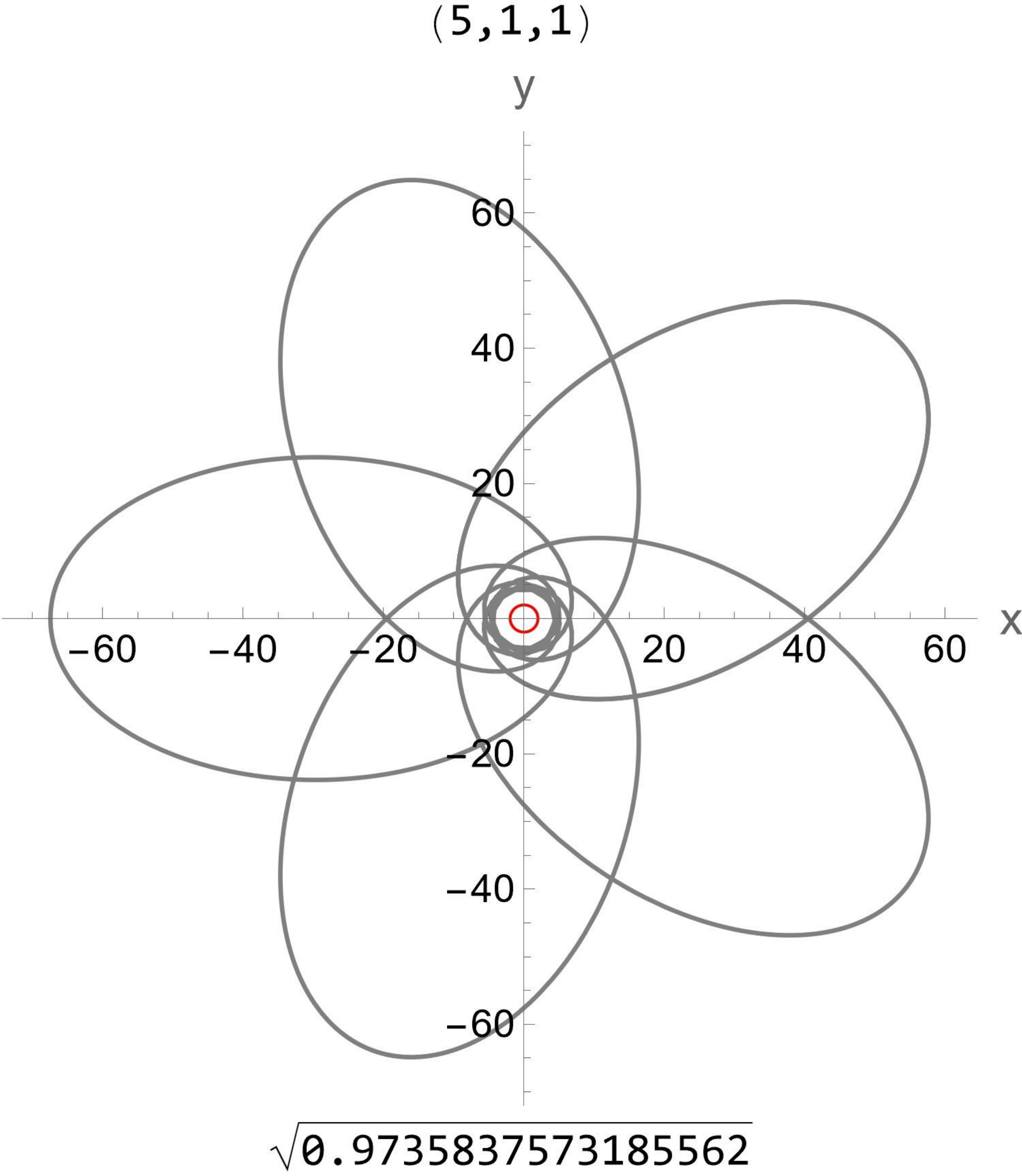}}
    \subfigure{
    \includegraphics[width=0.25\textwidth]
     {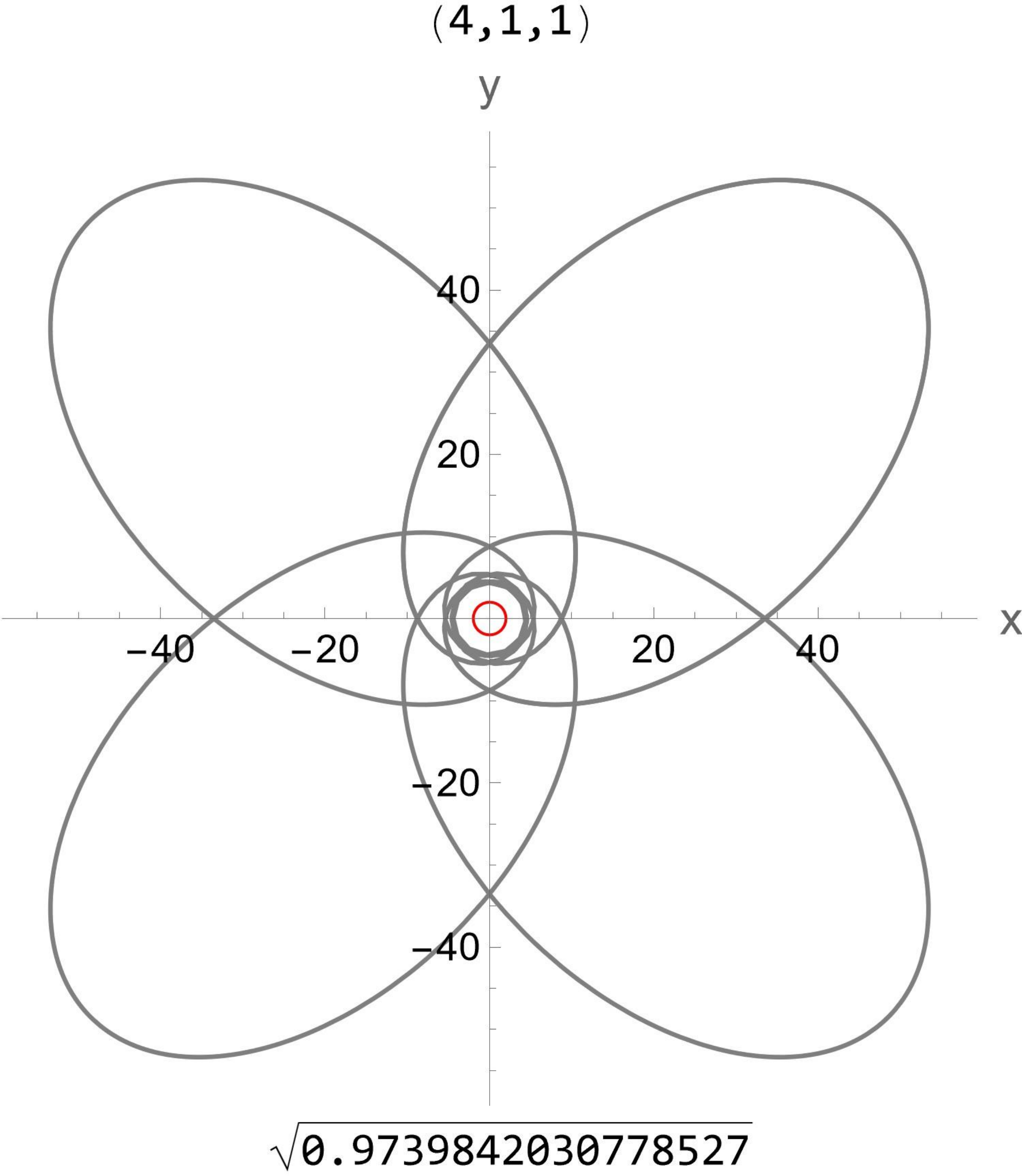}}
     
    \subfigure{
    \includegraphics[width=0.25\textwidth]
     {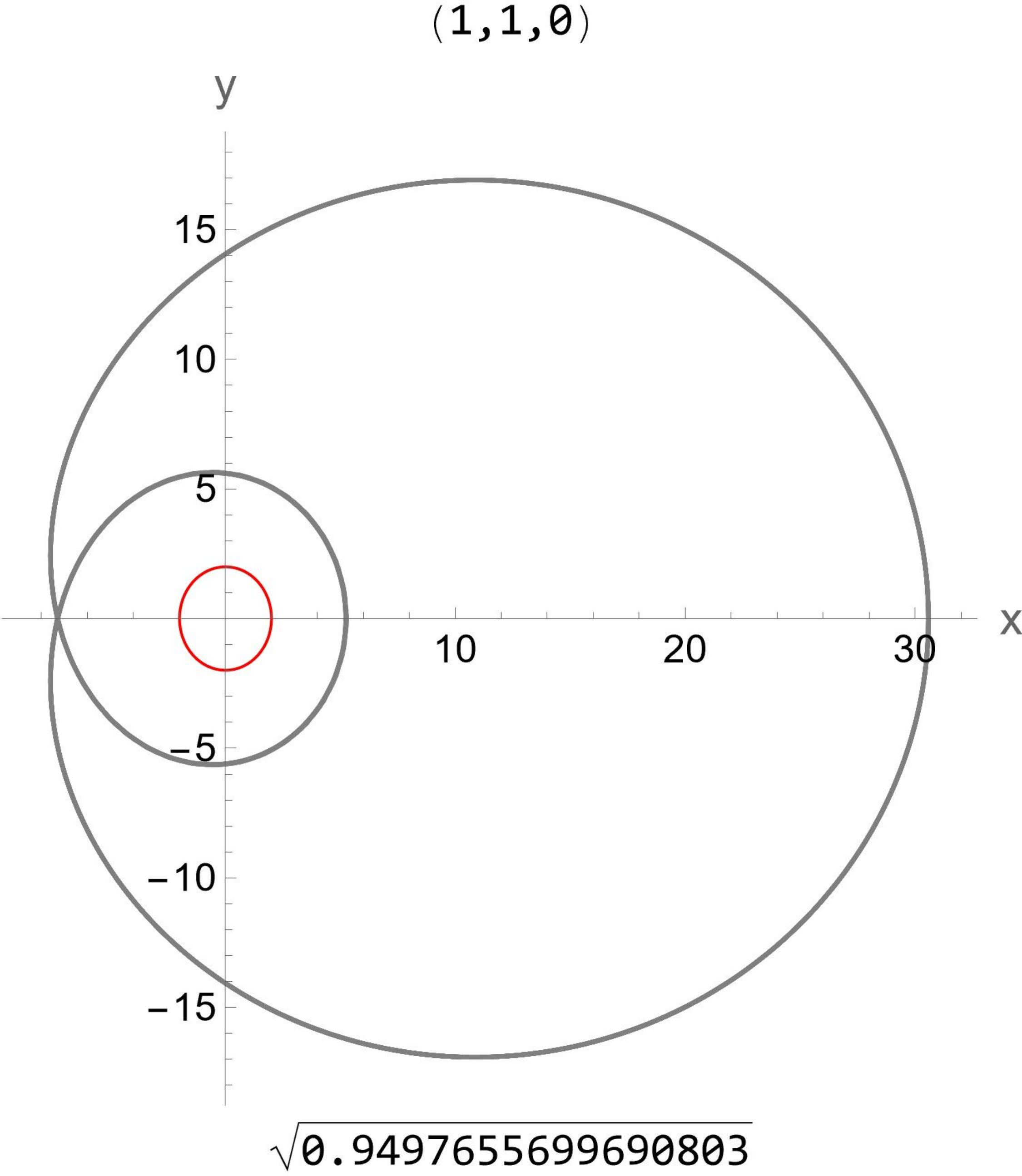}}
     \subfigure{
    \includegraphics[width=0.25\textwidth]
     {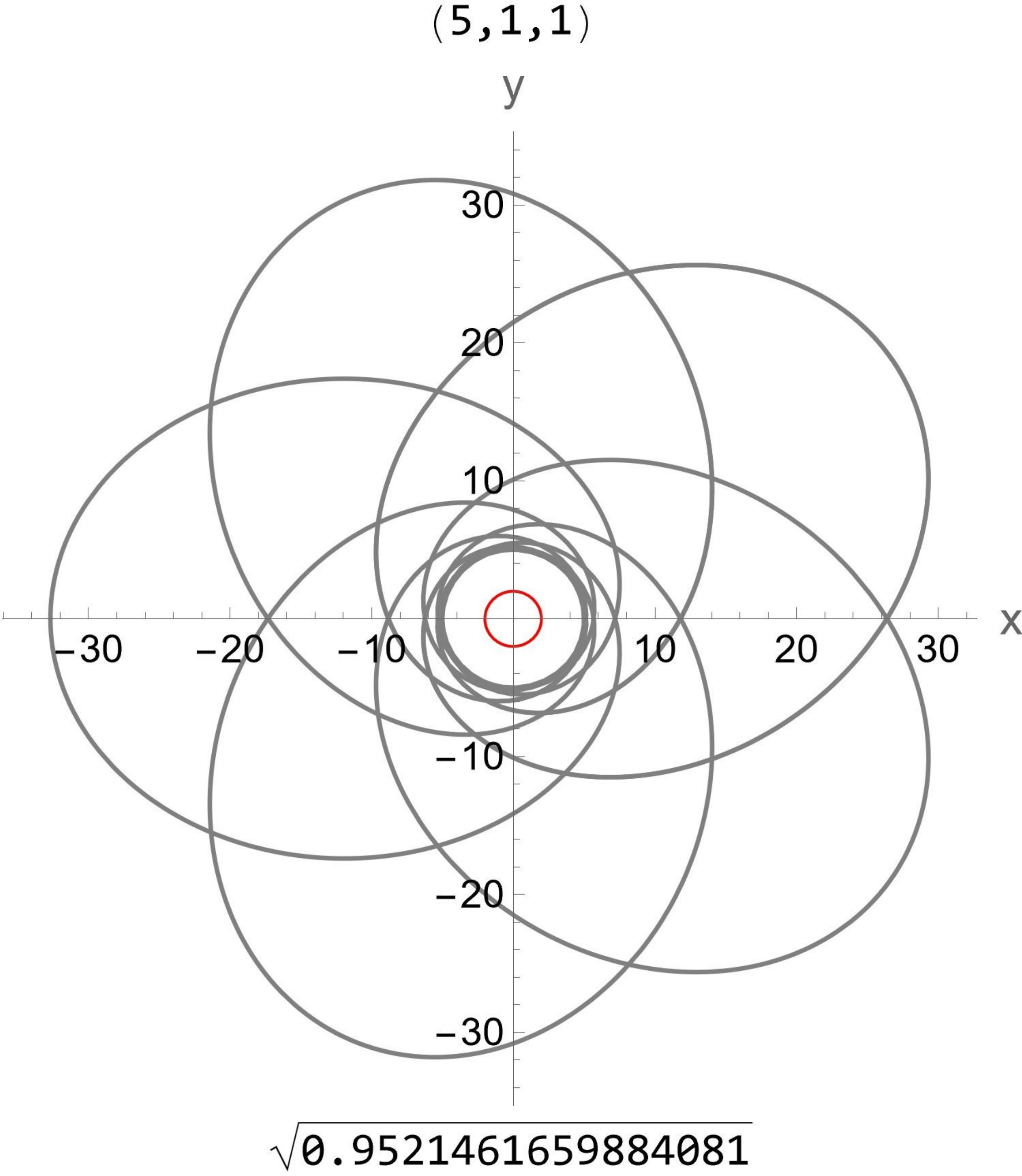}}
    \subfigure{
    \includegraphics[width=0.25\textwidth]
     {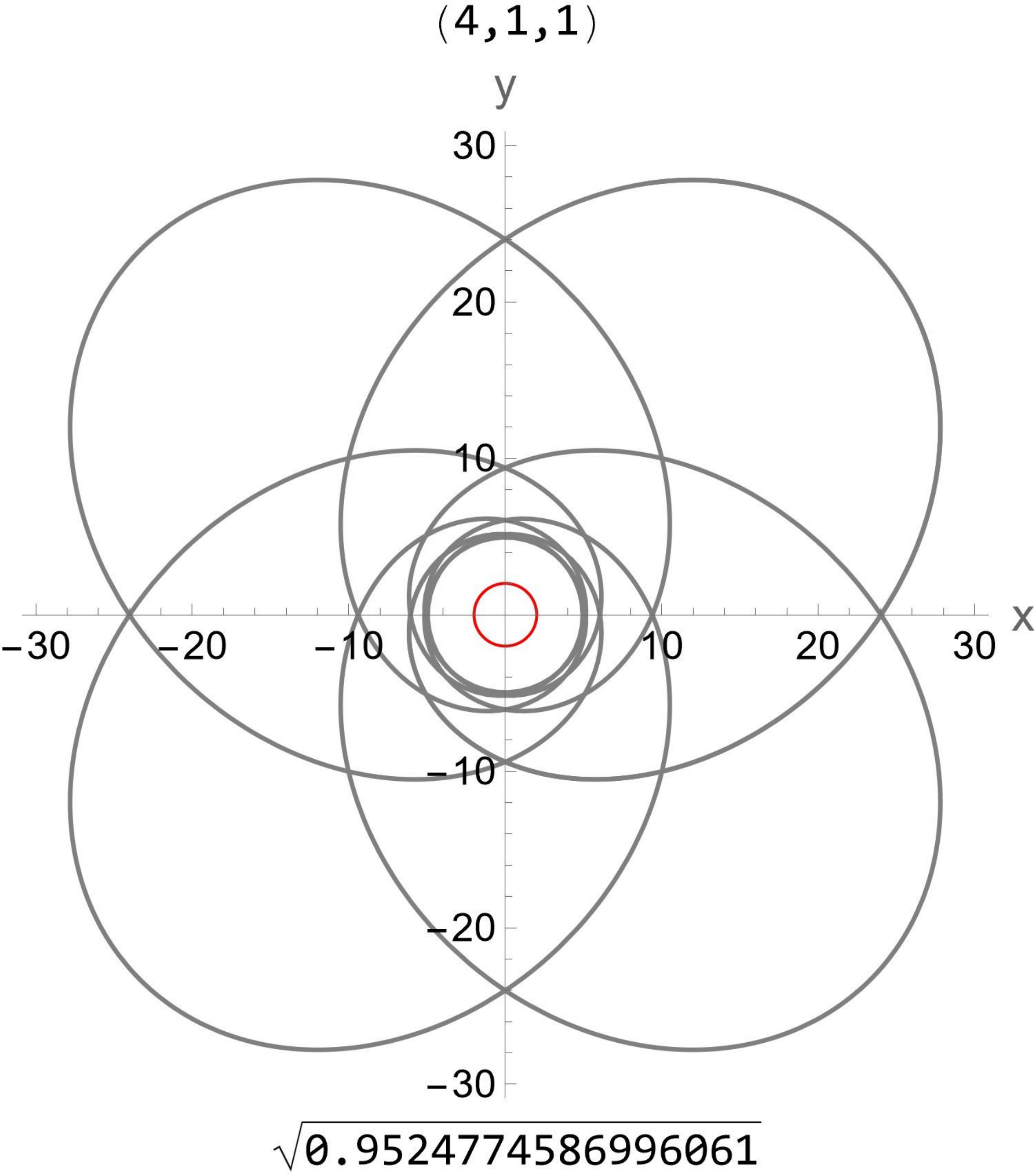}}

    \subfigure{
    \includegraphics[width=0.25\textwidth]
    {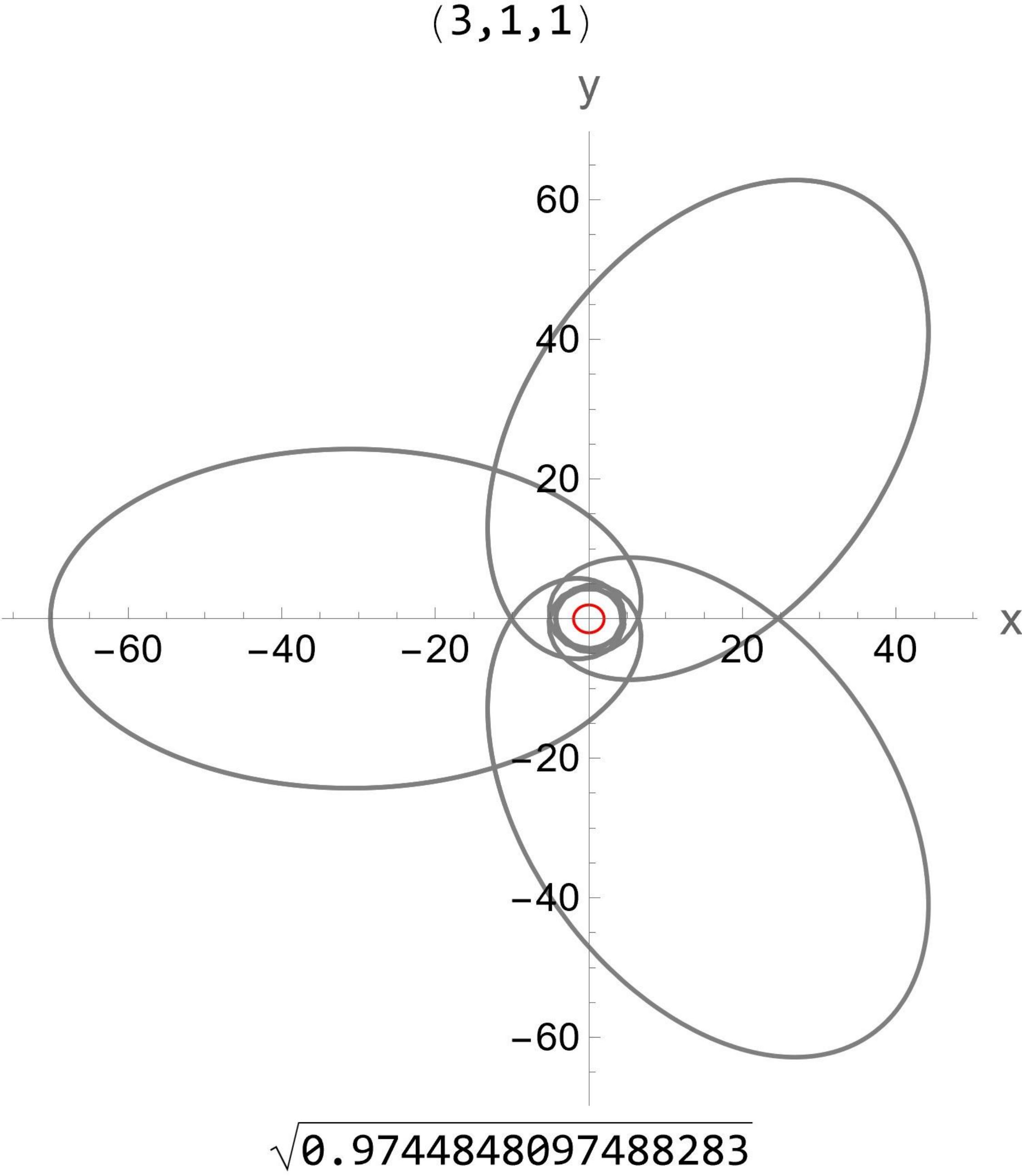}}
    \subfigure{
    \includegraphics[width=0.25\textwidth]
    {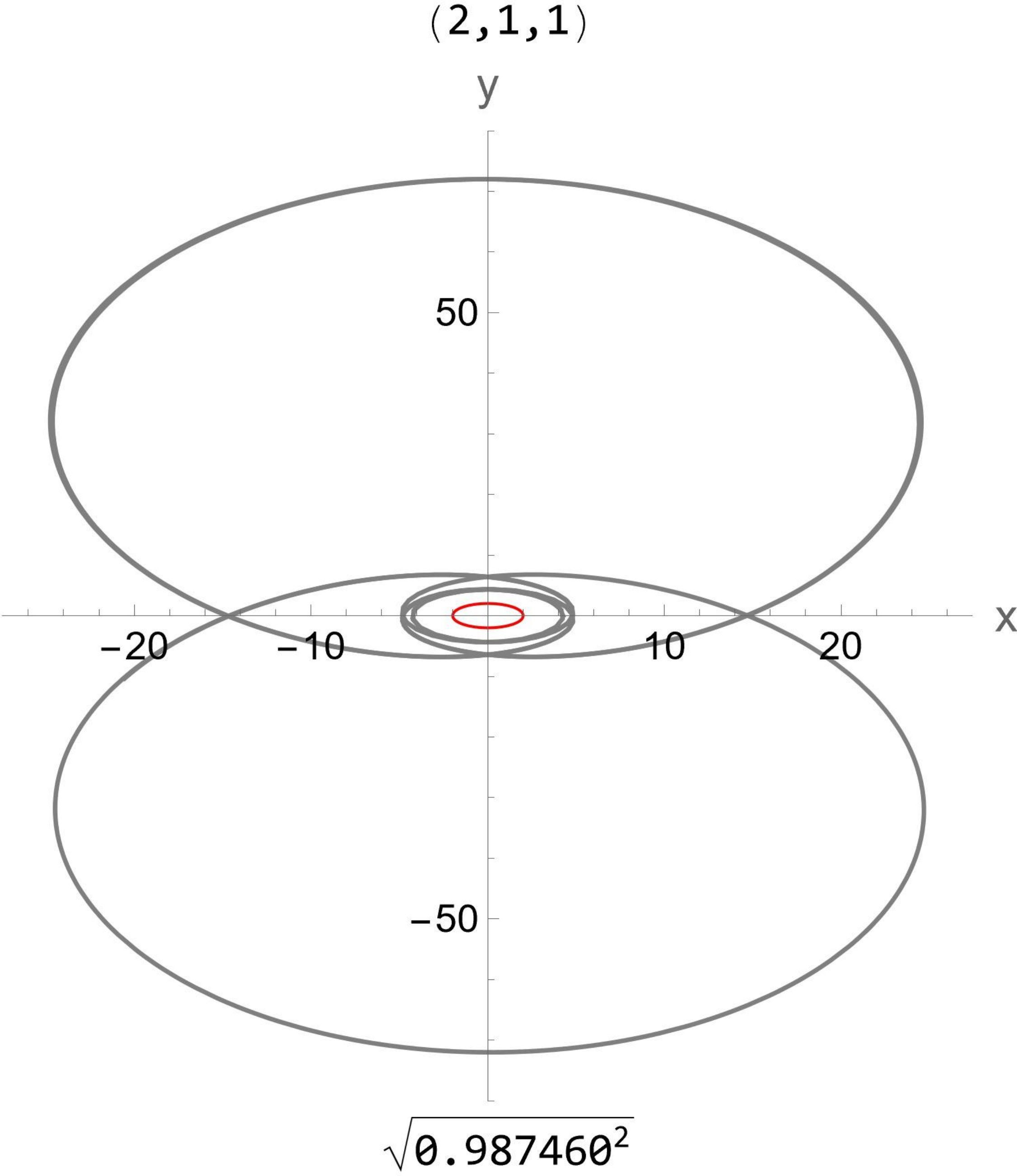}}
    
    \subfigure{
    \includegraphics[width=0.25\textwidth]
     {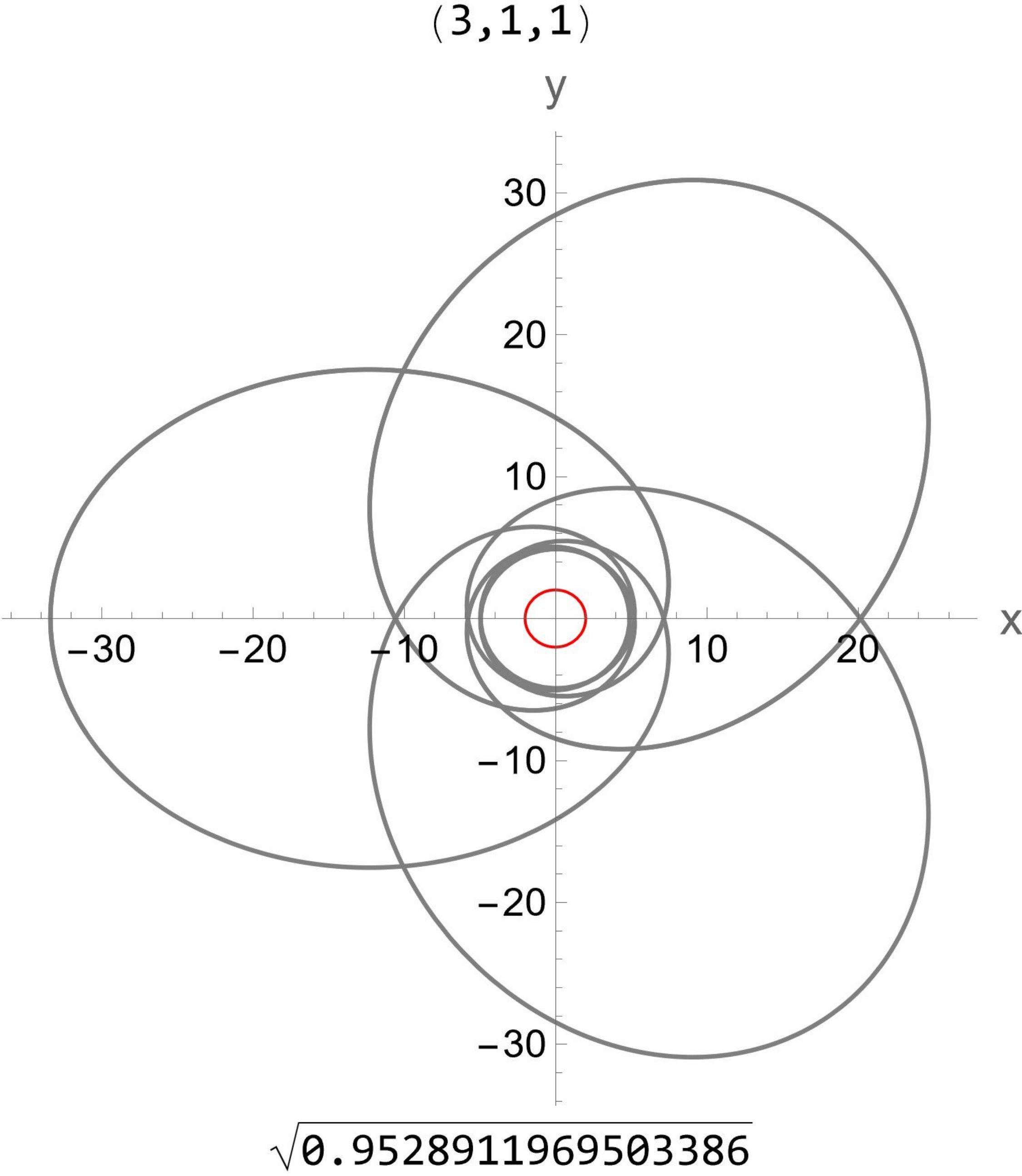}}
    \subfigure{
    \includegraphics[width=0.25\textwidth]
     {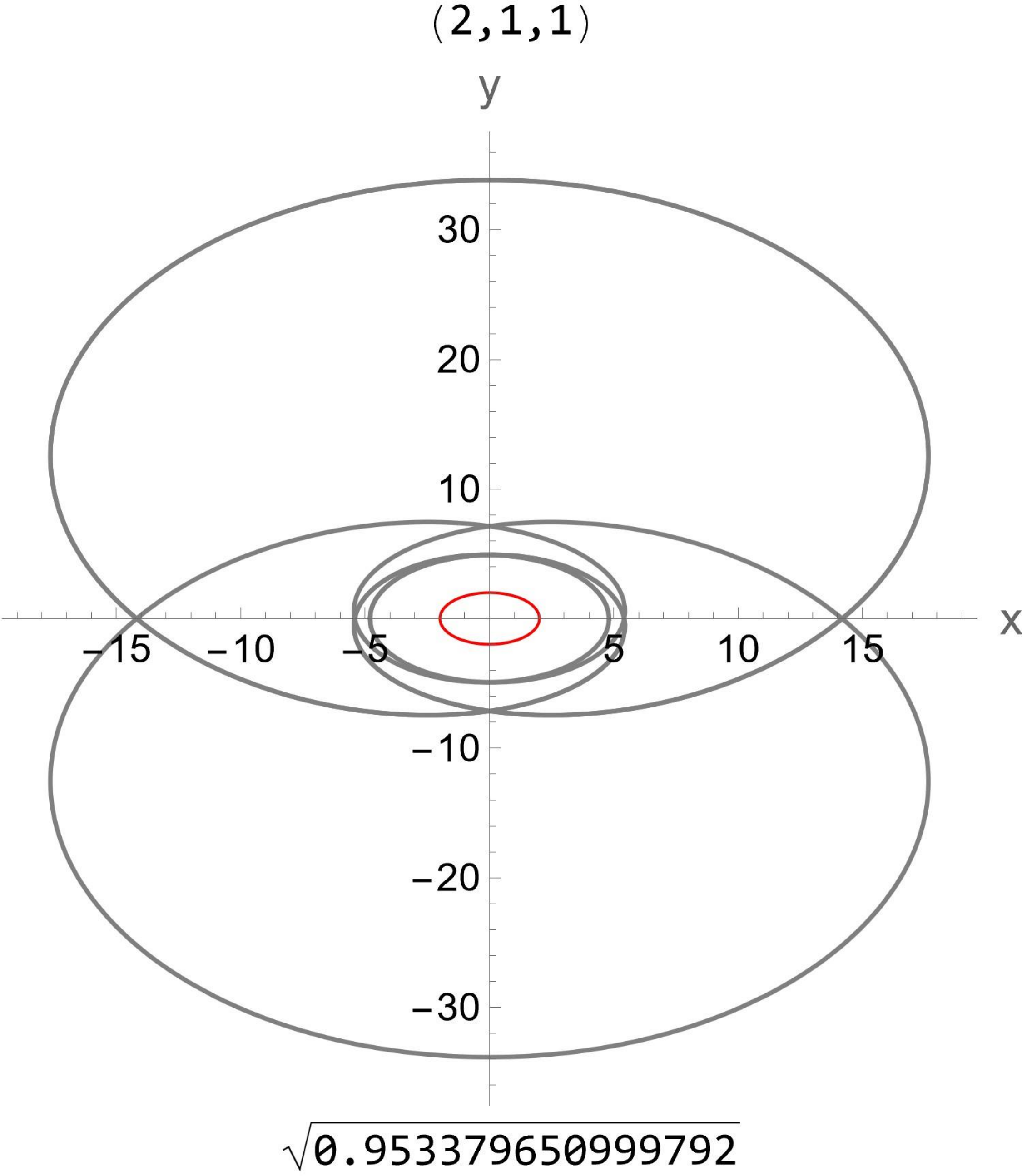}}  
    \caption{Different periodic orbits are presented with the fixed angular momentum $L_{z}=3.9$. The values of deformation parameter $\alpha$ are 0 and -1. The values of the triplet array for each orbit are plotted also.}
    \label{f5}
\end{figure*}

The deformation parameters of the other control group are $\alpha=0$ and $\alpha=-1$. Since when $\alpha=0$ the situation has degraded to Schwarzschild spacetime, in which the angular momentum should be larger than $\sqrt{12}$ and there will be periodic orbits, the angular momentum is fixed at a constant value of $\sqrt{15.21}$. Similarly, the integer is from 1 to 1.5 for each set of orbits. The trend of perihelion and the apastron changing with deformation parameter changes showed in Fig.\ref{f5} is similar to the previous group.

\section{Gravitational Radiation from periodic orbits}\label{sec5}
In this section, a preliminary research of the gravitational radiation emitted by periodic orbits is provided. The gravitational radiation emitted may contain information about the spacetime, so it is valuable for research.

To compute the waveform, a common technique is the adiabatic approximation. Compared to the time of evolution on the order of $10^6$ of proper time, changes of energy $E$ and the orbital angular momentum $L_{z}$ are negligible during periods on the order from $10$ to $10^3$. This means that $E$ and $L_{z}$ can be set as constants. The waveform of periodic orbits will be calculated later without considering the influence on the motion.

The numerical Kludge model will be introduced briefly primarily, since it captures many qualitative features of EMRIs\citep{Chua:2017ujo, Babak:2006uv}. It can even be available in self-force condition by including the particle dynamics. Although the self-consistency will degrade, it is still effective as an approximation to Teukolsky-based waveform in strong field\citep{Chua:2017ujo}. To compute the waveform, equations of motion \eqref{e4} need be solved numerically to construct trajectory in space of motion constants firstly. As mentioned before, the influence caused by radiation reaction can be ignored. Secondly, the curved-space coordinates of trajectory then will be associated artificially with coordinates in flat space. Using the standard quadrupole formula, the corresponding gravitational waveform is generated.

The symmetric trace-free (STF) mass quadrupole of the small object with mass $m$ is\citep{Thorne:1980ru}:
\begin{equation}
    I^{ij}=\int  x^{i}x^{j}T^{tt}(t,x^{i})\mathrm{d}^{3}x.
\end{equation}
The $tt$-component of the stress-energy tensor for the small object is
\begin{equation}
    T^{tt}(t,x^{i})=\mu\delta^{3}[x^{i}-Z^{i}(t)].
\end{equation}
with $Z$ being the trajectory. So in the extreme-mass-ratio limit, the mass quadrupole moment will be 
$I^{ij}=\mu x^{i}x^{j}$. By adopting existing methods, the curved-space coordinates can be projected onto the Cartesian coordinate system:
\begin{equation}
    \begin{split}
        &x=r\sin\theta \cos\varphi,\\
        &y=r\sin\theta \sin\varphi,\\
        &z=r\cos\theta.
    \end{split}
\end{equation}
Considering the weak-field situation, the spacetime metric can be written down as deviation from the flat spacetime as $g_{\mu\nu}=\eta_{\mu\nu}+h_{\mu\nu}$, in which $\eta_{\mu\nu}$ is the flat spacetime and $h_{\mu\nu}$ is the perturbation. The perturbation can be written as follow:
\begin{equation}
    h_{ij}=\frac{2m}{D_{L}}(a_{i}x_{j}+a_{j}x_{i}+2v_{i}v_{j}),
\end{equation}
where $D_{L}$, $v_{i}$ and $a_{i}$ denote the luminosity distance from the detector to the source, the velocity and acceleration of the small object, respectively. A detector-dependent Cartesian coordinate system $(X,Y,Z)$ connected with the source frame $(x,y,z)$ is adapted here. The basis vectors of $(X,Y,Z)$ in the original coordinate system $(x,y,z)$ are:
\begin{equation}
  \begin{split}
    &\mathbf{e}_{X}=[\cos\eta, -\sin\eta,0],\\
    &\mathbf{e}_{Y}=[\sin\iota \sin\eta,-\cos\iota \cos\eta, -\sin\iota],\\
    &\mathbf{e}_{Z}=[\sin\iota \sin\eta, -\sin\iota \cos\eta, \cos\iota].
  \end{split}
\end{equation}
where $\iota$ and $\eta$ represent the inclination angle and the longitude of the perihelion of the orbital plane, respectively\citep{Barack:2003fp}. Then the gravitational wave polarizations can be obtained as follows:
\begin{equation}
    \begin{split}
        &h_{+}^{ij}=\frac{1}{2}(\mathbf{e}_{X}^{i}\mathbf{e}_{X}^{j}-\mathbf{e}_{Y}^{i}\mathbf{e}_{Y}^{j}),\\
        &h_{\times}^{ij}=\frac{1}{2}(\mathbf{e}_{X}^{i}\mathbf{e}_{Y}^{j}+\mathbf{e}_{Y}^{i}\mathbf{e}_{X}^{j}).
    \end{split}
\end{equation}
Considering the transverse-traceless gauge, the concrete expressions of crucial components are given as\citep{Babak:2006uv}:
\begin{equation}
    \begin{split}
        &h_{\eta\eta}=h_{xx}\cos^{2}\eta-h_{xy}\sin2\eta+h_{yy}\sin^{2}\eta,\\
        &h_{\iota\iota}=\cos^{2}\iota[h_{xx}\sin^{2}\eta+h_{xy}\sin2\eta+h_{yy}\cos^{2}\eta]+\\
        &\ \ \ \ \ \ \ \ \ \ h_{zz}\sin^{2}\iota-\sin2\iota[h_{xz}\sin\eta+h_{yz}\cos\eta],\\
        &h_{\iota\eta}=\cos\iota[\frac{1}{2}h_{xx}\sin2\eta+h_{xy}\cos2\eta-\frac{1}{2}h_{yy}\sin2\eta]+\\
        &\ \ \ \ \ \ \ \ \ \ \sin\iota[h_{yz}\sin\eta-h_{xz}\cos\eta].
    \end{split}
\end{equation}
Combining all equations above, the polarizations can be expressed as follows:
\begin{equation}
    \begin{split}
        &h_{+}=\frac{h_{\eta\eta}-h_{\iota\iota}}{2},\\
        &h_{\times}=h_{\iota\eta}.
    \end{split}
\end{equation}

\begin{figure*}
    \centering
    \subfigure{
    \includegraphics[width=0.3\linewidth]
    {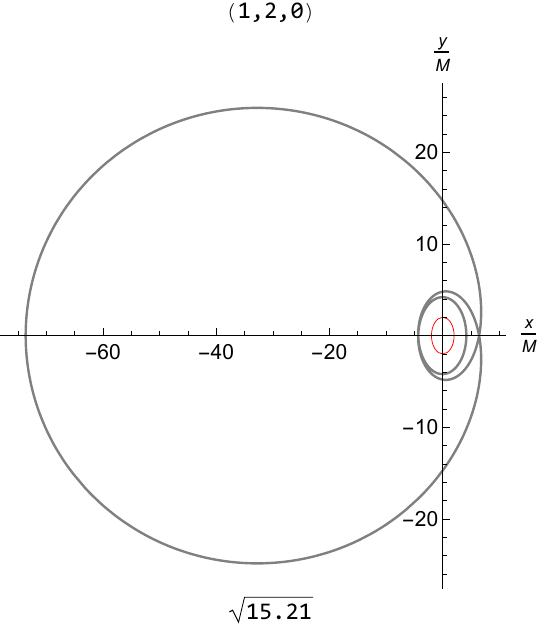}}
    \subfigure{
    \includegraphics[width=0.3\linewidth]
    {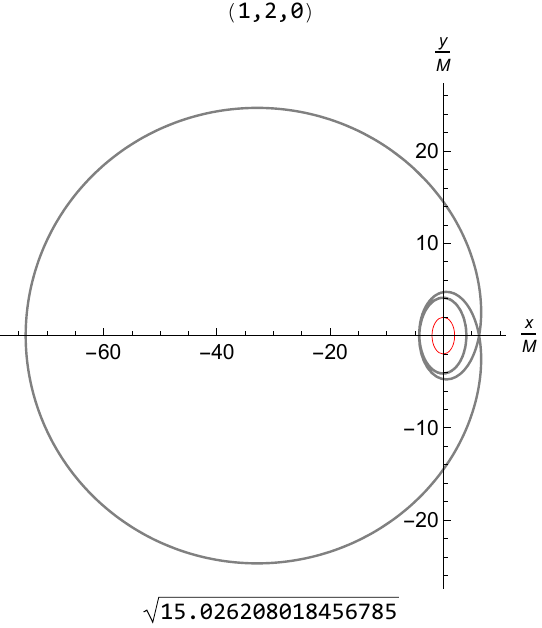}}
    \subfigure{
    \includegraphics[width=0.3\linewidth]
    {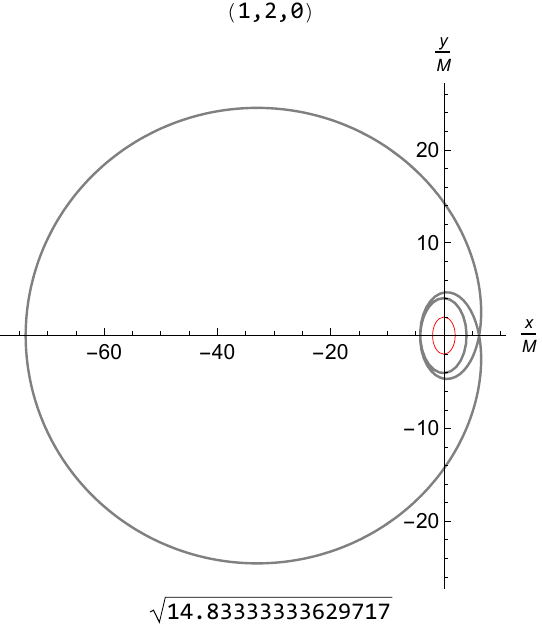}}
    \caption{Different periodic orbits with the energy $E$ fixed at $\sqrt{0.9755751102284231}$. Deformation parameter $\alpha=0$, $\alpha=0.2$, $\alpha=0.4$.}
    \label{f6}
\end{figure*}

\begin{figure*}
    \centering
    \subfigure{
    \includegraphics[width=1\linewidth]{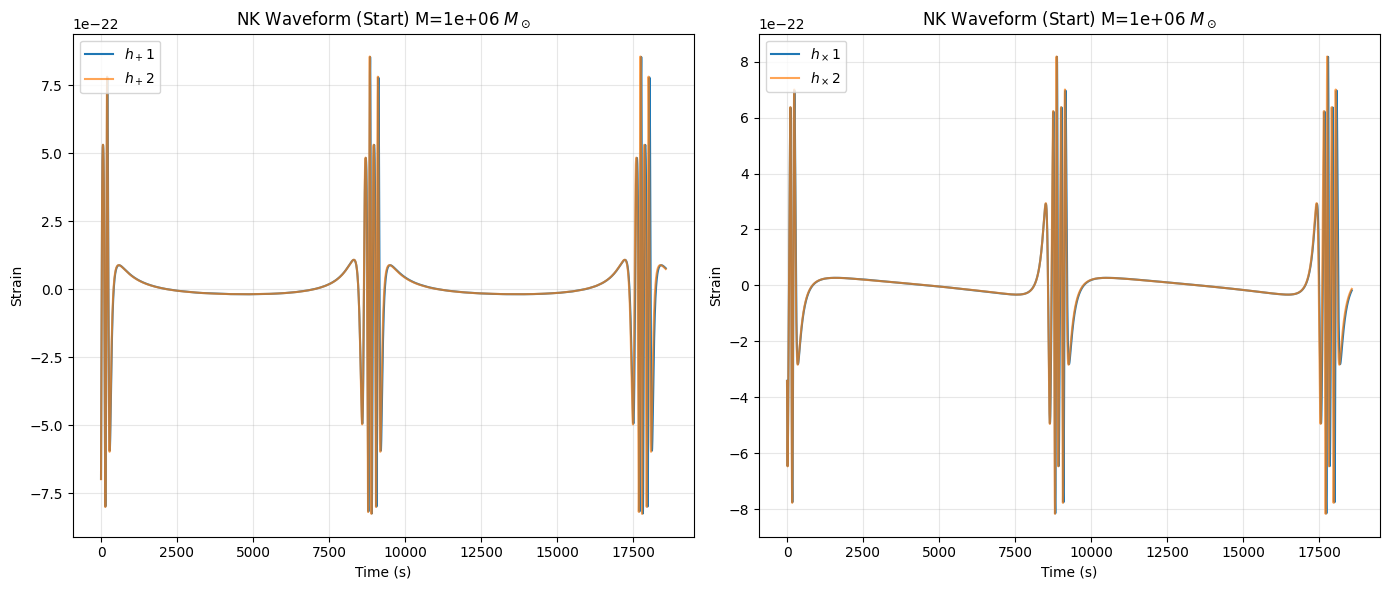}}
    \subfigure{
    \includegraphics[width=1\linewidth]{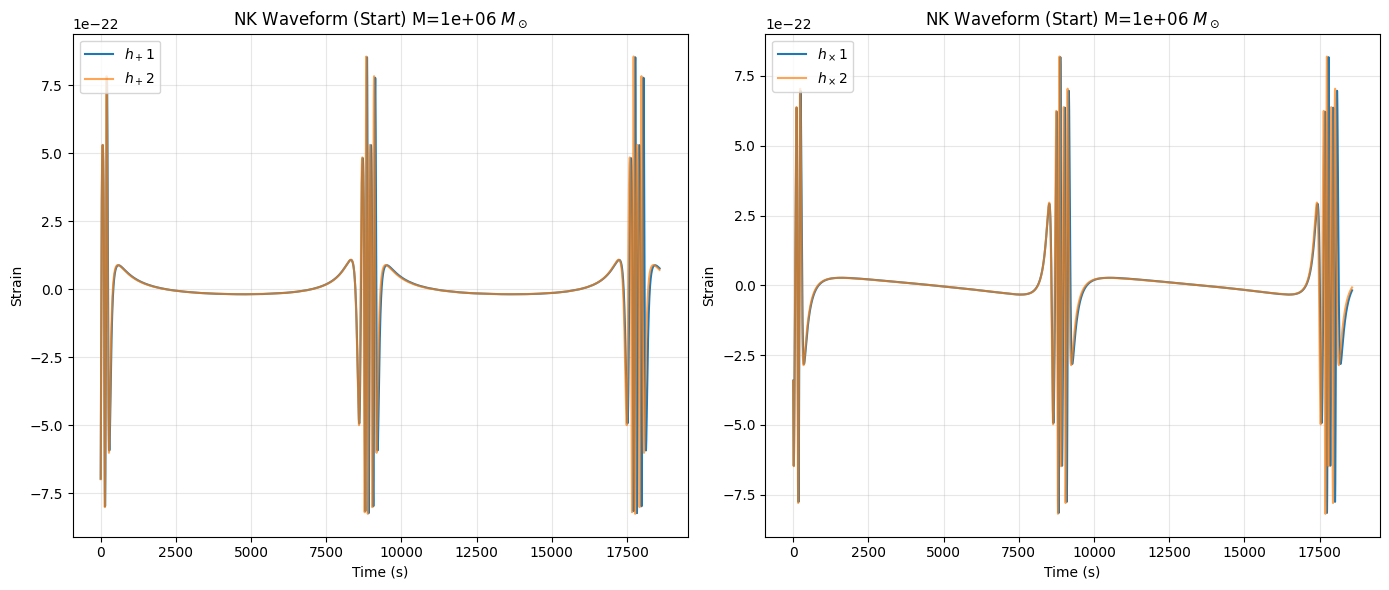}}
    \caption{Two polarization components of gravitational waveforms of periodic orbit from a test object with $m=10M_{\odot}$ around a supermassive black hole with $M = 10^6M_{\odot}$. Energy $E$ is fixed as $\sqrt{0.9755751102284231}$. The above figure shows the difference between $\alpha=0$ and $\alpha=0.2$. The below figure shows the difference between $\alpha=0$ and $\alpha=0.4$.}
    \label{f7}
\end{figure*}

In order to show the influence of the deformation parameter on the gravitational waveforms of different periodic orbits, we consider a small compact object with mass $m=10M_{\odot}$ around a supermassive black hole with mass $M=10^{6}M_{\odot}$ at the luminosity distance $D_{\text{L}}=1\text{Gpc}$. Take a simple case, the inclination angle $\iota$ is set as $\pi/2$ and the longitude of perihelion is fixed at 0. Moreover, the polarization angle is also set as 0 for simplicity\citep{Babak:2006uv}. To show contrast, we fix the energy $E$ at $\sqrt{0.9755751102284231}$, change the deviation parameter $\alpha$, and also use method of bisection to get $L_{z}$. When $\alpha$ is large enough, the perihelion is too small, as shown in Fig.\ref{f6}. So numerical kludge model will not work. And then the waveforms are represented by two polarization components $h_{+}$ and $h_{\times}$ to show their characteristics in Fig.\ref{f7}. We list three parameter conditions and their waveform contrast here.

In Figs \ref{f6} and \ref{f7}, a typical kind of periodic orbits is plotted with the triplet $(1,2,0)$ and their $h_{+}$ and $h_{\times}$ modes with different deformation parameters. It shows that a clear relationship exists between gravitational waveforms and orbit. The giant leaf results in very long smooth phases. 
And there are rapid oscillations which align with the zoom and whirl behaviors. Moreover, it is clear that in comparison with the Schwarzschild case the parameter influences the phase of the waveforms mainly and the waveform amplitudes in a slight way when parameter increases. To clearly quantitatively illustrate the differences in gravitational waveforms with different parameters, a useful characteristic is the mismatch between two waveforms\citep{Allen:2019vcl, Qiao:2024gfb, Lai:2025zfm}. The specific calculation formula is given as\citep{Pompili:2025rhz}:
\begin{equation}
    \mathcal{M}=1 - \max_{\varphi, t_c} \frac{(h_1 \mid h_2)}{\sqrt{(h_1 \mid h_1)(h_2 \mid h_2)}}
\end{equation}
While $h$ is defined as $h=h_{+}-ih_{\times}$. To conduct a preliminary analysis of the differences, we calculated the mismatch according to the Schwarzschild condition with $\alpha$ from 0 to 0.38 and concluded the trend below in Fig.\ref{f8}. To show this trend more comprehensively, we display mismatch of three different periodic orbits, with the triplet $(1,2,0)$, $(2,1,1)$ and $(3,1,1)$, respectively. We display the smooth curve with the marked original data points in the figure.:
\begin{figure}
    \centering
    \includegraphics[width=1\linewidth]{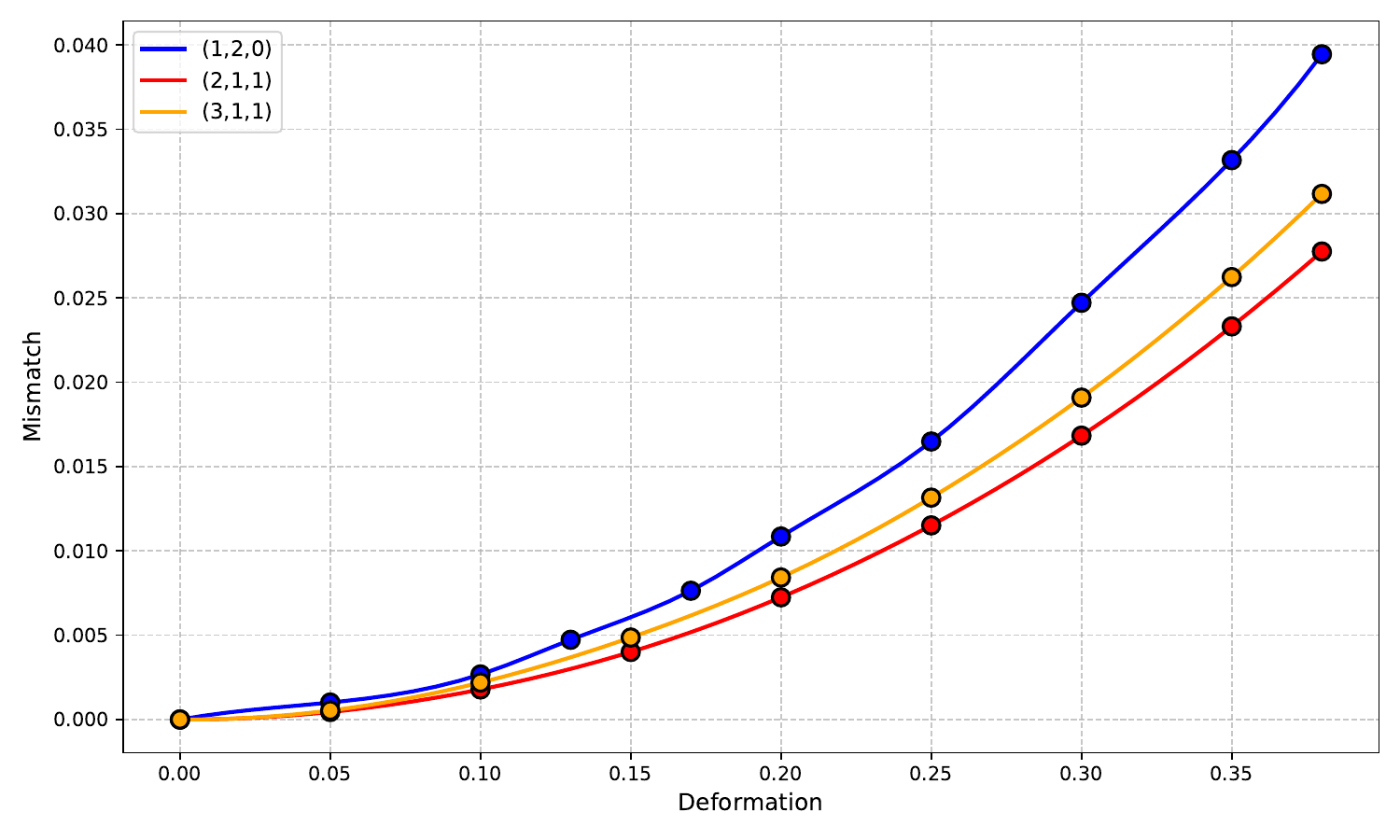}
    \caption{Mismatch based on the Schwarzschild case with the parameter $\alpha$ from 0 to 0.38. The blue, red and yellow line shows the mismatch of orbits with the triplet $(1,2,0)$, $(2,1,1)$ and $(3,1,1)$, respectively. The bolded points represent the original data.}
    \label{f8}
\end{figure}

Although not shown in Fig. \ref{f8}
, we find that the mismatch exceeds $0.05$ for $\alpha \geq 0.4$. The mismatch increases with the deformation parameter, as expected for progressively larger departures from the Schwarzschild situation. We also find that the mismatch varies across different periodic orbits, indicating an orbit-dependent sensitivity to the deformation. It is noteworthy that with the same number of $w$ and $v$, the mismatch of the orbit with three branches is greater than that of the orbit with two branches, which may be attributed to the accumulation of waveform differences induced by the number of orbital branches. While the deformation induces only modest changes in the waveform amplitude and phase, the mismatch provides a preliminary means of discriminating among these cases. A definitive assessment, however, requires a dedicated analysis of detectability and measurability, following and extending the methodology of  Ref.\citep{Speeney:2022ryg, Barausse:2006vt, Moore:2017lxy, Cardoso:2021wlq, Cardoso:2022whc, Destounis:2022obl, Gliorio:2025cbh, Datta:2025ruh, Rahman:2023sof}. Overall, these results suggest a plausible pathway toward probing subtle physical features in future gravitational wave detection.

\section{Conclusions and Discussions}\label{sec6}
In this paper, we have studied the geodesic motion of a small compact object orbiting a supermassive non-rotating DSK black hole. To determine the possible motion region of the small object, we started with derivation of the radial effective potential of photons and massive particles. We carefully solved the system of equations of circular orbit for photon and got the relationship between the parameter $\alpha$ and the radius.
It is very interesting that there is a fixed horizon radius and the photon orbits will shrink when $\alpha$ increases until reach horizon and disappear. The orbital angular momentum, meanwhile, will decrease with the same trend. However, there are some unexpected results when we applied similar methods to massive particles. For one thing, there are two kinds of circular orbits, which are MBOs and MSCOs. For the other, these two kinds of orbits will branch out when deformation parameter $\alpha$ reaches a specific value. Two branches will meet together when $\alpha$ reaches the second specific value. After that, there are no circular orbits when $\alpha$ becomes bigger. These are unique characteristics different from Schwarzschild spacetime and they make the motion region different as well. Small objects can move in two separated regions which are called inner region and outer region, and motions in outer region are similar to those in the Schwarzschild spacetime. Then we use the taxonomy of periodic orbit to distinguish different orbits by the triplet array $(z, w, v)$. We then plot different periodic orbits in the inner and outer region, respectively. The motion in the inner region, however, usually has large value of the triplet array. This is quite different from the Schwarzschild spacetime. And the periodic orbits in the outer region are similar to the Schwarzschild spacetime. Our research indicates that the energy of the periodic orbits increases while the orbital angular momentum decreases with the growth of parameter $\alpha$.

We also preliminarily considered the gravitational waveform from periodic orbits in the outer region. The system is composed of a small compact object with mass $m=10M_{\odot}$ around a supermassive black hole with mass $M=10^{6}M_{\odot}$. We carefully calculated the polarizations $h_+$ and $h_\times$ with three different values of $\alpha$ and put them together to demonstrate the difference. The waveform signals showed a clear relationship with topological properties of periodic orbits. 
Leaves generate the phases smoothly while zoom and whirl behaviors generate rapid oscillations. These changes show the transition from weak field to strong field vividly. With fixed energy, there are some subtle differences in phases and amplitude. To make it clearer, we calculated the mismatch of waveforms compared with the Schwarzschild spacetime. The value of mismatch goes up in total with the growth of $\alpha$. This shows the possibility of distinguishing between different gravitational waveform signals.

The work presented here reveals some distinct features of geodesics from the Schwarzschild black hole. Nevertheless, there are some intriguing problems to address in the future. For example, orbital resonances can be taken into further investigation\citep{Flanagan:2010cd}. 
Moreover, the possible mechanism in the inner region and the disappearance of the photon ring are also worth considering. Last but not the least, while research for geodesics in DSK metric under general conditions is challenging, qualitative analysis can be taken into consideration. We leave these topics to future work.

\section*{Acknowledgements}
We thank Hongbao Zhang for helpful discussion. This research work is supported in part by the National Key Research and Development Program of China (grant number 2020YFC2201300), and the National Natural Science Foundation of China (grant numbers 12035016, 12375058 and 12361141825).

\bibliographystyle{elsarticle-harv} 
\bibliography{ref}






\end{document}